\documentclass[twocolumn,preprintnumbers,amsmath,amssymb,nofootinbib
,superscriptaddress]{revtex4}
\usepackage{hyperref}
\usepackage{graphicx}
\usepackage{color}
\usepackage{xcolor}
\usepackage{comment}
\usepackage{lmodern}

\newcommand{\backo}{\!\!\!\!\!\!\!\!\!\!}

\newcommand{\smebo}{\eea  }
\newcommand{\smbbo}{\bea  && }
\newcommand{\smallebo}{ \normalsize \eea  \normalsize}
\newcommand{\smallbbo}{\small  \bea  &&   }
\newcommand{\bl}{\biggl(}
\newcommand{\br}{\biggr)}
\newcommand{\Tr}{\makebox{Tr}}
\newcommand{\vvx}{\vec{x}}
\newcommand{\vvy}{\vec{y}}
\newcommand{\vvr}{\vec{r}}
\newcommand{\vvX}{\vec{X}}
\newcommand{\vvq}{\vec{q}}
\newcommand{\vvp}{\vec{p}}
 
\newcommand{\be}{\begin{equation}}  
\newcommand{\ee}{\end{equation}}  
\newcommand{\bea}{\begin{eqnarray}}   
\newcommand{\eea}{\end{eqnarray}}  
\newcommand{\ba}{\begin{array}}  
\newcommand{\ea}{\end{array}}

\usepackage{diagbox}

\definecolor{rossoCP3}{cmyk}{0,.88,.77,.40}
\definecolor{blueRef}{rgb}{0.2,0.2,0.6}
\definecolor{blue}{rgb}{0,0.396,0.741}

\hypersetup{
	colorlinks, 
	bookmarksopen, 
	bookmarksnumbered,
	citecolor=blueRef, 		
	linkcolor=rossoCP3,	
	urlcolor=rossoCP3,			
}

\newskip\humongous \humongous=0pt plus 1000pt minus 1000pt

\newif\ifdtup

\def\oldreffmt#1{\rlap{[#1]} \hbox to 2\parindent{}}

\def\figfmt#1{\rlap{Figure {#1}} \hbox to 1in{}}  
  
\def\Tr{\mathop{\rm Tr}}

\def\slash#1{#1\!\!\!/\!\,\,} 	
\def\beq{\begin{equation}}  
\def\eeq{\end{equation}}  
\def\bea{\begin{eqnarray}}  
\def\eea{\end{eqnarray}}  
\def\half{\frac{1}{2}}  
  
\def\bq{\begin{quote}}  
\def\eq{\end{quote}}

\def\half{\frac{1}{2}}    
\newcommand{\nonumbo}{\nonumber \\ && }
  
\def \gta {\mathrel{\vcenter  
     {\hbox{$>$}\nointerlineskip\hbox{$\sim$}}}}   

\relax  

\newcommand{\U}{\mathrm{U}}

\newdimen\tdim  
\tdim=\unitlength  
\def\bar{\overline}

\begin{document}

\preprint{FERMILAB-PUB-24-0720-T}

\title{A New-Old Approach to Composite Scalars 
\\
with Chiral Fermion Constituents 
}
\author{Christopher T. Hill}
\email{chill35@wisc.edu}
\affiliation{Fermi National Accelerator Laboratory,
P. O. Box 500, Batavia, IL 60510, USA}
\affiliation{Department of Physics, University of Wisconsin-Madison, Madison, WI, 53706
}

\begin{abstract}
We develop a dynamical, Lorentz invariant theory of composite scalars 
in configuration space consisting of chiral fermions, 
interacting by the perturbative exchange of a massive ``gluon'' of coupling $g_0$ and mass $M_0^2$ (the coloron model).
The formalism is inspired by, but goes beyond, old ideas of Yukawa and the Nambu-Jona-Lasinio (NJL) model.
It yields a non-pointlike internal wave-function of the bound state, $\phi(r)$,
which satisfies a Schr\"odinger-Klein-Gordon (SKG) equation
with eigenvalue $\mu^2$. For super-critical coupling, $g_0 >g_{0c}$, we have $\mu^2< 0$ leading to spontaneous symmetry breaking. 
The binding of chiral fermions is semiclassical, {\em
not loop-level as in NJL }. 
The mass scale is determined by the interaction as in NJL. 
We mainly focus on the short-distance, large $M_0^2$ limit, yielding an NJL pointlike
interaction, but the bound state internal wave-function,
 $\phi(\vvr)$, remains spatially extended and dilutes $\phi(0)$. This leads to
 power-law suppression of the induced Yukawa and
quartic couplings and requires radically less fine-tuning of a hierarchy than does the NJL model.  
We include a discussion of loop corrections of the theory. A realistic top condensation model
appears possible.
\end{abstract}

\maketitle
 
\date{\today}

\email{chill35@wisc.edu}


\section{Introduction }


There are many lines of investigation
of composite models of bound states
of chiral fermions that can be 
treated more-or-less analytically,
 including
\cite{Feynman,BagModels,CJT,Yukawa,NJL} to name a few. 
However, the Nambu--Jona-Lasinio model (NJL) \cite{NJL} stands out as a
concise and useful Lorentz invariant description of dynamical bound 
states of relativistic chiral fermions
in quantum field theory.  

The NJL model is fairly easy
to implement, tying an underlying chirally invariant fermionic action 
to composite scalars and dynamical symmetry breaking.
It operates  at the quantum loop level, ${\cal{O}}(\hbar)$,
and is most readily solved 
by using the renormalization group (RG) \cite{Wilson,KogutWilson,BHL,CTH},
and shares features with BCS superconductivity 
in the large $N_c$ (fermion color) limit \cite{BCS}.
A large literature exists of successful applications to the chiral
dynamics of QCD, \cite{Bijnens,NJLReview,bardeenhill}, and it provided
the first 
models of a composite Brout-Englert-Higgs (BEH) boson 
and related phenomena \cite{NSD,BHL,Yama,Topcolor,HMTT}.

There are, however, fundamental physical limitations of the NJL model:
(1) the NJL model is an effective pointlike 4-fermion interaction
associated with a ``large'' mass scale $M_0$; 
(2) the resulting bound states emerge as {\em pointlike} fields with mass $\mu^2< M_0^2$;
(3) in the NJL model 
the binding mechanism is entirely driven by quantum loop effects, while 
we see in nature that binding readily occurs
semiclassically without quantum loops, such as the Hydrogen atom.

 In the case of the Hydrogen atom, before turning on the Coulomb interaction, there
are open scattering states involving free protons and electrons.  As the interaction
is adiabatically turned on the lowest energy scattering states flow to become the bound states, 
while most scattering states remain unbound. 
The dynamics is governed by the non-relativisitic, semiclassical
Schr\"odinger equation \cite{Schrodinger} in an extended potential, leading
to normalizable, yet spatially extended,  wave-functions on the scale $(\alpha m_e)^{-1}$.
The atom is described naturally in a configuration space picture.
Quantum loop effects (such as the Lamb shift) are higher order corrections to this mostly semiclassical phenomenon.

In the NJL model the picture is substantially different. There is no semiclassical binding
producing an extended bound state. Rather, the bound state is described by a {\em local, or pointlike,} effective field, $\Phi(x)$, with its constituents arising in quantum loops.  The loops integrate out the 
fermions from the large mass scale of
the interaction, $M_0$, down to an IR cut-off $\mu$
(e.g., $M_0\sim 1$ GeV and $\mu\sim f_\pi \sim 100$ MeV in QCD). The discussion can be formulated
in momentum space treated in the
large $N_{color}$ limit, where bound states appear as poles in the S-matrix upon summing
leading-$N_c$ fermion loop diagrams.  With a large hierarchy $M_0/\mu >\!\!> 1$ there
are also large logarithms, and the solution is best handled by constructing
the effective Lagrangian and using
the renormalization group (RG). We summarize this procedure
for the NJL model in Appendix \ref{NJLreview}.

When we say ``pointlike'' we are referring to the explicit local form of $\Phi(x)$.
In the Wilsonian RG picture 
the field at any scale $m<M_0$ can be viewed as
{\em effectively} pointlike, on distance scales $r>m^{-1}$.
However, $\Phi(x)$ has only the minimal dynamical degrees of freedom of a pointlike
field.
The NJL model is well suited to study the chiral dynamics 
of light quarks in QCD where ``integrating out the fermions'' mimics confinement
and there are no 
free scattering states of massless chiral quarks at large distances.
Any analogy with the Hydrogen atom does not apply.

In the case of chiral dynamics in a {\em non-confining theory}, however,
one may ask what has become of semiclassical
binding of chiral fermion pairs?  
Indeed, here there arises an apparent logical conflict in the NJL model if  we  tune the coupling constant near its  ``critical value.''  
In this case  the pointlike 
bound state becomes nearly massless, and
at the critical coupling we have a conformal effective low energy
theory. In configuration space, conformality  
implies a nearly scale invariant extended internal wave-function, $\phi(r) \propto 1/r$,
(the limit of  $\phi(r)\propto e^{-|\mu|r}/r$ for mass $|\mu|<\!\!< M_0$). This
extends to large distances, well beyond the range of the potential $\sim  M_0^{-1}$. Hence
we would expect that the near--critical
bound state must always be an extended object, even if the interaction scale, $M_0^{-1}$,
is a very short distance scale.  We will argue that the NJL model,
lacking an internal wave function, becomes misleading in this limit, and essentially fails.

To address and expand upon these issues, 
we presently explore a Lorentz invariant formulation in configuration space 
by introducing a fully dynamical color singlet composite bilocal
 field, $\Phi(x,y) \sim \bar{\psi}_R(x)\psi_L(y)$. 
This formulation contains the requisite internal
wave-function with its own kinetic term. This is a ``new-old'' 
approach to compositeness as it hearkens back to Schr\"odinger \cite{Schrodinger} 
and relativistic generalizations of Yukawa \cite{Yukawa}. 
However, we still lean heavily on the NJL model to provide intuition
while extending it to a  non-pointlike, bilocal field theory. 
Introducing   $\Phi(x,y)$ has the immediate advantage of bosonizing the composite theory (we can then mostly bypass issues of Dirac operators until
we do loops). Including $\phi(r)$ into the structure of the theory 
introduces new degrees of freedom and leads to interesting consequences.

Consider a pair of massless particles of 4-momenta $p_1$ and $p_2$, 
where $p_1^2=p_2^2=0$,
and a two-body ``scattering state'' consisting of plane waves,
$\Phi(x,y) \sim \exp(ip_1x+ip_2y)$.
We introduce ``barycentric coordinates,'' $X=(x+y)/2$, 
 and $r=(x-y)/2$, and corresponding momenta $P=p_1+p_2$, $Q=p_1-p_2$. Hence
 we can write $\Phi =\chi(X)\phi(r)$
where $\chi(X) = \exp(iPX)$ governs the ``center of mass motion'' 
and $\phi(r)=\exp(iQr)$ is the ``internal wave-function''
of the system.  This factorization of $\Phi$ will be our ansatz
for bound states.

Since $P_\mu Q^\mu=0$, then in the rest frame we have
$P=(P_0,\vec{0})$ and $Q=(0,\vvq)$. 
Hence, the dependence upon  
the relative time $r^0$ drops out, and
the internal wave-function is static  $\phi(\vvr)$.
We construct the Hamiltonian for $\phi(\vec{r})$
including interactions.  The extremalization of
this Hamiltonian leads to the ``Schr\"odinger-Klein-Gordon'' (SKG) equation for $\phi(r)$, with eigenvalues 
$\mu^2$ for the (mass)$^2$ of bound state solutions.

The most natural
UV completion of the interaction of the NJL model is the ``coloron model'' \cite{Bijnens, Topcolor,Simmons}.
The coloron is essentially a massive gluon, associated with an $SU(N_c)$ gauge theory broken to
a global $SU(N_c)$, featuring an adjoint representation of massive gauge bosons of mass $M_0$.  Integrating
out the colorons (and removing relative time)
yields a perturbative static Yukawa potential 
in the rest frame, 
\smallbbo
\backo
\label{one}
-g_0^2N_c M_0 V_0(2r) \;\;\makebox{where,}\;\; V_0(2r)=-\frac{e^{-2M_0r}}{8\pi r}.
\smallebo
This is a non-pointlike 
generalization of the NJL interaction. 
The potential is enhanced by a factor of $N_c$ colors, in analogy to the $N_{Cooper}$ (Cooper pairs)
enhancement of the Fr\"ohlich interaction in the 
BCS theory of superconductivity \cite{BCS}, and arises here
by the color singlet normalization of $\Phi(x,y)$.  $M_0$ is the defining
mass scale of the theory.

The bound state forms semiclassically.
Even in a short-distance potential, (e.g., the large $M_0>\!\!|\mu|$ limit
of eq.(\ref{one})) this leads
to a spatially extended internal bound state wave-function, $\phi(r)$.
As $g_0^2$ approaches a critical value, $g_0^2\rightarrow g_{0c}^2 $
where $\mu^2\rightarrow 0$, then $\phi(r)\rightarrow 1/r$ is conformal.
For supercritical coupling $g_0^2N_c$
the eigenvalue of the SKG equation, $\mu^2$,  {\em is always negative (tachyonic)}  and
a chiral vacuum instability and 
dynamical spontaneous symmetry breaking will therefore occur classically.

The
most interesting result in this formalism is  a significant
wave-function spreading of $\phi(r)$, where the large distance part  
dominates the normalization of $\phi(r)$ and dilutes $\phi(0)$.
There is an induced Yukawa coupling
of the bound state to free fermions, $g_Y\propto \phi(0)$, 
which gives the {\em  unbound}
fermions their dynamical mass, and is
therefore suppressed
by the infrared dilution, as  $\phi(0)\sim \sqrt{|\mu|/M_0}$. Hence, $g_Y$ is then significantly smaller
than the Yukawa coupling obtained in the NJL model, which runs only logarithmically.

Moreover, this has the direct consequence of reducing the sensitivity of the eigenvalue
to the near critical value of the coupling constant,
 significantly reducing the degree of fine-tuning needed to have
a hierarchy between $M_0^2$ and $\mu^2$.
We also compute loops in this formalism which
depend upon $g_Y$, consistent with the dilution effect. The loops 
can generate
 the quartic coupling, $\lambda \propto g_Y^4 \ln(M_0/\mu)$
 which is necessary to stabilize the vacuum
 in a tachyonic, supercritical solution.  Again, the dilution
 effect leads to a smaller value of $\lambda$ than expected in the NJL model.

This ``infrared dilution'' effect drastically contrasts the results from the NJL model where the 
RG running of the Yukawa coupling into the IR is logarithmic and comparatively slow.
If, for example, this applies to the BEH boson, as in
a top quark condensation scheme \cite{Topcolor}, then 
it can naturally arise as an extended composite object
on the scale of the electroweak VEV, $v = 175$ GeV with $M_0\sim 5$ to $ 10$ TeV.  
The dilution of $\phi(0)$ effectively isolates the strong interaction at $M_0$ from the
low energy physics and eliminates the need for drastic fine-tuning in these
models, yielding fine-tuning at the few $\%$ level.
The scale symmetry near criticality acts here as the
custodial symmetry for the emergent low mass scale physics.

We give a lightning review of the familiar NJL model with pointlike
potential treated in Appendix \ref{NJLreview}, and discuss key aspects of it in the next section. 
We then formulate bilocal fields, 
and construct a model of an extended potential
that arises from a single massive gluon exchange (the coloron model),
 suitably Fierz rearranged.
The formalism leads to the SKG equation.  
We then analyze the induced Yukawa
coupling, and induced quartic coupling (the latter from loops),
and consider properties of the solutions to the SKG equation
by variational methods.
We show that the critical coupling in the semiclassical Yukawa potential case is 
almost identical to the critical coupling of the NJL model!  The wave function spreading 
and power-law suppression of the Yukawa and quartic coupling emerges and
the fermion loop renormalizations are consistent with the suppression
and perturbative.
Appendices contain detailed extensions of the discussion.
Some precursory work to the present appears in
\cite{bilocal} 


\section{Bilocal Theory of Composite Scalars}

\subsection{Key Aspects of the Nambu--Jona-Lasinio Model \label{keyNJL} }

 Here we touch on key elements relevant to
formulating a bilocal description of bound states.
A detailed review of the RG approach to the NJL model is presented
in Appendix \ref{NJLreview}. 

The NJL model \cite{NJL}
assumes chiral fermions,  with $N_c$ ``colors.''
A chirally invariant  $U(1)_{L}\times U(1)_{R}$
action takes the form:
\smallbbo
\label{0NJL1}
S_{NJL} 
=\!\int\! d^4x \;\bl i[\bar{\psi}_L(x)\slash{\partial}\psi_{L}(x)]
+ i[\bar{\psi}_R(x)\slash{\partial}\psi_{R}(x)]
\nonumber \\ &&
\qquad \qquad
+\;
\frac{g_0^2}{M_0^2}
[\bar{\psi}_L(x)\psi_{R}(x)]\;[\bar{\psi}_R(x)\psi_{L}(x)]
\br.
\smallebo
Here the notation $[..]$ implies color singlet combination.

We rewrite eq.(\ref{0NJL1}) in an equivalent form
by introducing a {\em local auxiliary
field} $\Phi(x)$:
\smallbbo
\label{0NJL2}
\backo
S_{NJL}
=\!\int\! d^4x \;\bl i[\bar{\psi}_L(x)\slash{\partial}\psi_{L}(x)]+ i[\bar{\psi}_R(x)\slash{\partial}\psi_{R}(x)]
\nonumbo
- M_0^2\Phi^\dagger(x) \Phi(x) + g_0[\bar{\psi}_L(x)\psi_{R}(x)]\Phi(x)+h.c.  \br.
\smallebo
The resulting ``equation of motion'' for $\Phi$ is:
\smallbbo
\label{0NJL222}
 M_0^2\Phi(x) = g_0[\bar{\psi}_R(x)\psi_{L}(x)]
\smallebo\normalsize 
Substituting the $\Phi(x)$ equation  back into eq.(\ref{0NJL2}) 
reproduces  the 4-fermion interaction of eq.(\ref{0NJL1}).
We emphasize that $\Phi(x)$ is a complex, local, or ``pointlike,'' field.
Generalizing the relation of eq.(\ref{0NJL222}) is the starting point for our
 bilocal field theory.
 
Note that the induced Yukawa coupling, $g_0$, in eq.(\ref{0NJL2}) is the same coupling as
appears in the interaction of eq.(\ref{0NJL1}).  This implies that strong coupling required
to induce chiral symmetry breaking in the NJL model will translate into a strong Yukawa coupling.
In our bilocal theory this will not be the case.

Following Wilson, \cite{Wilson},
we view eq.(\ref{0NJL2}) as the action defined at the high energy 
(short-distance) scale $m \sim M_0$.
We then integrate out the fermions to obtain the effective action for the composite 
field $\Phi$ at a mass scale of the low
energy physics, $\mu <\!\!<M_0$ \cite{BHL,CTH}.  
In leading $N_c$ fermion loop approximation this yields:
\smallbbo
\label{0NJL3}
\backo
S_{\mu}
=\!\int\! d^4x \;\bl i[\bar{\psi}_L\slash{\partial}\psi_{L}]+ i[\bar{\psi}_R\slash{\partial}\psi_{R}]
+
Z\partial_\mu \Phi^\dagger\partial^\mu \Phi
\nonumbo
\backo
-\mu^2\Phi^\dagger \Phi - \frac{\lambda }{2}(\Phi^\dagger \Phi)^2 + (g_0[\bar{\psi}_L\psi_{R}]\Phi(x)+h.c. )
\br.
\smallebo\normalsize 
where,
\smallbbo
\label{0NJL4}
\backo
\mu^2 = M_0^{2}\!-\!\frac{g_0^{2}N_c}{8\pi ^{2}} M_0^{2},
\nonumbo
\backo
Z=\frac{g_0^{2}N_c}{8\pi ^{2}}\ln(M_0/\mu), \;\;\;
\lambda=\frac{g_0^{4}N_c}{4\pi ^{2}}\ln( M_0/\mu).
 \smallebo\normalsize
Here $M_0^2$ is the UV loop momentum 
cut-off, and we obtain induced kinetic and quartic interaction terms.
The one-loop result can be improved by using the full renormalization group (RG) \cite{BHL,CTH}.

It is a feature of the NJL model that apparently all
of the fermions in the scale range $M_0$ to $\mu$ are integrated out to comprise 
a ``tightly bound state.'' As seen in the formalism of Cornwall, Jackiw and Tomboulis (CJT)
\cite{CJT}, however,
one is integrating out the {\em quantum fluctuations}
of the fermions, and we could introduce classical fermion sources 
to describe the residual unbound scattering state fermions.\footnote{ We are unaware of any
formal treatment of the NJL model in the CJT formalism, which would be of interest, but it must dovetail
with the RG formalism; in our own attempt we found this challenging.}
Generally, in the RG approach, unbound fermions in the scale range $M_0$ to $\mu$
are simply assumed to remain in the action.

Note the behavior of the composite scalar boson mass, 
$\mu^2$, of eq.(\ref{0NJL4}) due to
the $ -{N_{c}g^{2}M_0^{2}}/{8\pi ^{2}}$   with
 UV cut-off  $M_0^{2}$. 
The NJL model therefore has a critical
value of its coupling defined by the vanishing of $\mu^2$,
\smallbbo
\frac{g_{0c}^{2}N_c }{8\pi ^{2}}=1
\smallebo\normalsize

We can renormalize, $\Phi\rightarrow \sqrt{Z}^{-1}\Phi$, 
to obtain the full effective Lagrangian.  
The notable feature here is that the renormalized Yukawa and quartic couplings
evolve logarithmically in the RG running mass $m$:
\smallbbo
\label{0NJL30}
\backo
\backo
g^2_Y \!=\! \frac{g_0^2}{Z}\!=\!\frac{8\pi^2}{N_c\ln(M_0/m)},\;\;
\lambda_r\!=\!\frac{\lambda}{Z^2}\! =\!\frac{16\pi ^{2}}{N_c\ln( M_0/m)}.
 \smallebo\normalsize
These are the solutions to the RG equations in the large $N_c$ limit,
\cite{CTH}, neglecting other interactions.  This indicates
that the couplings approach a Landau pole as $m\rightarrow M_0$,
and we use the full RG equations with this boundary condition
to obtain the low energy solution in the NJL model \cite{BHL}.
The evolution into the IR is then gradual, 
in particular, $g_Y$ approaches an IR fixed point value \cite{PR}.
In our present scheme, the value of $g_Y$ at $m=M_0$ 
is determined by $\phi(0)$, is finite
and power-law suppressed, hence the IR value can be below the fixed point.
 
For  super-critical coupling, $g_0^2>g^2_c$,
we see that $\mu^2<0$ and there will be a chiral vacuum instability.
The effective action, with the induced quartic  $\sim \lambda_r(\Phi^\dagger\Phi)^2$ term, 
is then the usual sombrero potential.
The chiral symmetry is spontaneously broken and  $\Phi$ acquires a VEV, and produces 
Nambu--Goldstone bosons and a Higgs boson.
The additional predictions of the NJL model are discussed Appendix \ref{NJLreview}.


\subsection{Semiclassical and Non-Point-like Generalization of the NJL Model}

 Consider a semiclassical approach to binding in a non-confining theory.\footnote{
``Semiclassical'' implies $\hbar=0$  (no loops). The leading term in
the $\hbar$ expansion of a quantum field theory is the sum over all tree-diagrams
which is equivalent to a
(Fredholm expansion) of a classical nonlinear field theory, 
 expressed in terms of frequencies, $\omega$, and wave vectors, $\vec{k}$.
Terms of order $\hbar^N$ 
in the expansion contain $N$-loops.} 
In the limit of shutting off the interaction a bound state
is just a two-body scattering state, such as a product of a free electron 
and free proton wave-functions in the case of Hydrogen.  For chiral fermions this can
be described by a bilocal field $\Phi^A_B(x,y)=\bar\psi{}^A_R(x) {\psi}_{LB}(y)|_b$, 
where $(A,B)$ arbitrary color and flavor indices and $b$ denotes the subset
of fermions that will comprise the bound state.
We also have the remaining unbound
free fermionic scattering states, that will remain free after the
interaction is turned on
(denoted by subscript $f$). 

A superposition of the bound and free fermions can be written as,
\footnote{\label{previous} This
can be viewed as a matrix element of  the operator $\bar{\psi}_R(x) {\psi}_{L}(y)$ sandwiched between
the vacuum, $\langle 0 |$ and a superposition of quantum states, 
  $|\Phi^A_B\rangle + |\bar{\psi}_R^A, \psi_{LB}\rangle$,
where  $|\Phi^A_B\rangle$ is a coherent state.}
\smallbbo
 \label{Phi01}
 \bar\psi{}^A_R(x) {\psi}_{LB}(y)= \bar\psi{}^A_R(x) {\psi}_{LB}(y)_{f} + M^2\Phi^{A}_{B}(x,y)
 \smallebo\normalsize 
 The components are orthogonal,
 \smallbbo
 \label{0Phi01}    
 \int_{xy}\bar\psi{}^A_R(x) {\psi}_{LB}(y)_{f}\Phi^\dagger{}^{A'}_{B'}(x,y)=0
 \smallebo\normalsize 
 Note we have implicitly defined $\Phi$ as a mass dimension-1 field, like a scalar,
 and the mass prefactor, $M^2$, should be viewed as part of the wave-function and
 will be elaborated below.
 
 Eq.(\ref{Phi01}) has a formal similarity to the factorized auxiliary field
 expression in eq.(\ref{0NJL222}). However, 
here $\Phi^{A}_{B}(x,y)$ is a distinct physical bilocal field and 
 its kinetic term is not induced by loops, and will have a free field kinetic term.
 While the NJL auxiliary field is not present in the NJL theory when the interaction
 is turned off,  the ingredients of the bound state are indeed 
 present in the semi-classical scheme in the absence of the interaction.
 
 We'll presently restrict ourselves to a single flavor, hence a $U(1)_L\times U(1)_R$
 flavor symmetry, and $(A,B)\rightarrow (i,j)$
 are $SU(N_c)$ color indices (this can be readily extended to $G_L\times G_R$ flavor groups as in
 Appendix (\ref{currentsapp})).

In the coloron model, of the next section, we will see that only the color singlet field
forms a bound state of a pair of chiral fermions.
With $SU(N_c)$ color indices, $(i,j)$, the field
$\Phi^{i}_{j}(X,r)$ is a complex matrix that transforms as a product
of $SU(N_c)$ representations,  $\bar{N}_c\times  {N}_c$, and
therefore decomposes into a singlet plus an adjoint representation.
We designate the color singlet bilocal field
as $\Phi^0$ and conventionally normalize it as,
\smallbbo
\label{singlet}
{\Phi}^{i}_{j}(x,y) =  \frac{1}{\sqrt{N_c}}\delta^i_j\Phi^0(x,y)
\smallebo\normalsize
This normalization allows canonically normalized kinetic terms,
$\Tr[\partial \Phi^\dagger \partial \Phi]
= \partial \Phi^0{}^\dagger\partial\Phi^0$.
Note $\Tr \Phi = {\Phi}^{i}_{i}(x,y) =  {\sqrt{N_c}}\Phi^0(x,y)$.\footnote{
In the NJL model, where $\Phi$ has no bare kinetic term,
the color normalization arises automatically 
at loop level together with the kinetic term.} 
Since only the color singlet binds, we can rewrite eq.(\ref{Phi01})
containing free fields and the color singlet bound state of eq.(\ref{singlet}),
 \smallbbo
\label{act2}
\backo\!\!\!\!\!\!\! \bar{\psi}^{i}_L(x)\psi_{jR}(y)
\rightarrow  \bar{\psi}^{i}_L(x)\psi_{jR}(y)_{f}
+ M^2\frac{\delta{^i_j}}{\sqrt{N_c}} \Phi^{0}(x,y).
 \smallebo\normalsize

\subsection{The Coloron Model $\label{3}$}

The pointlike NJL model can be viewed as the limit of a physical theory
with a bilocal interaction. The primary example  is the   ``coloron model'' 
\cite{Topcolor,NSD,Simmons} (the coloron idea stems from
QCD applications of NJL \cite{Bijnens} and the top condensation theory \cite{Topcolor} ).
The coloron is a perturbative, massive gauge boson, analogue
of the gluon, arising in a 
local $SU(N_c)$ gauge theory broken to a global $SU(N_c)$.
We integrate out the massive coloron and keep the single particle exchange
potential to define the model. This yields a bilocal current-current form.

\smallbbo
\label{TC0}
\backo
\;S'\!=\!-
{g_0^2}\!\!\int_{xy}\!\!\! 
 [\bar{\psi}_{L}\!(x)\!\gamma_\mu T^A \psi_{L}\!(x)]D^{\mu\nu}(x-y)
[\bar{\psi}{}_{R}(y)\! \gamma_\nu T^A \psi_{R}(y)]
\nonumbo
\smallebo\normalsize
where $T^A=T_i^{Aj}$ are generators of $SU(N_c)$, and color singlet 
combinations indicated in brackets $[...]$.
We adopt a notational convenience 
for integrals as defined in Appendix {\ref{SumNote}},

The coloron propagator in Feynman gauge is:
\smallbbo
\label{propagator}
D_{\mu\nu}(x-y)= g_{\mu\nu} D_F(x-y)
\nonumbo
D_F(x-y) =- \int\frac{1}{q^2-M_0^2}e^{iq(x-y)}\frac{d^4q}{(2\pi)^4}.
\smallebo\normalsize
A Fierz rearrangement of the interaction to leading order in $1/N_c$ 
leads to a potential:
\smallbbo
\label{coloronexchange}
\backo
S'=g_0^2\!\!\int_{xy}\!\! \; [\bar{\psi}_L(x)\psi_{R}(y)] D_F(x-y)[\bar{\psi}_R(y)\psi_{L}(x)],
\smallebo\normalsize 
(see Appendix \ref{spinor} for details, of the Fierz rearrangement).
The above displayed term is the most attractive channel and leading
in large $N_c$.

Hence, we
replace the pointlike 4-fermion interaction with 
the non-pointlike  $S'$ of eq.(\ref{coloronexchange}).
Note that if we suppress the $q^2$ term in the denominator
of eq.(\ref{propagator}) we have,
 \smallbbo
\label{4NJL}
{D}_F(x-y)\rightarrow  \frac{1}{M_0^2}\delta^4(x-y),
 \smallebo\normalsize
and  we recover the pointlike NJL model interaction.

Now,
substitute eq.(\ref{act2})  into eq.(\ref{coloronexchange}) to obtain,\footnote{
This
can equivalently be viewed as the matrix element of eq.(\ref{coloronexchange}) 
sandwiched between the quantum states, 
  $|\Phi^0\rangle+ |\bar{\psi}_R, \psi_L\rangle$.
}
\smallbbo 
\label{5NJL}
S'
\rightarrow
g_0^2\!\!\int_{xy}\!\! [\bar{\psi}_L(x)\psi_{R}(y)]_f D_F(x-y) [\bar{\psi}_R(y)\psi_{L}(x)]_{f}
\nonumbo 
+{ g^2_0\sqrt{N_c}}M^2\!\!\int_{xy}\!\![\bar{\psi}_L(x)\psi_{R}(y)]_{f}  D_F(x-y)\;\Phi^0(x,y) {+h.c.}
\nonumbo
+ { g_0^2N_c} M^4\!\!\int_{xy}\!\! \; \Phi^0{}^\dagger(x,y)\; D_F(x-y)\;\Phi^0(x,y)
\smallebo\normalsize
\normalsize
The leading term $S'$ of eq.(\ref{5NJL}) is the unbound  4-fermion
scattering interaction 
and has the structure of the NJL interaction in the limit of eq.(\ref{4NJL}) and
identifies $g_0$ as the analogue of the NJL coupling constant.
The second term $\sim g^2_0\sqrt{N_c}[\psi^\dagger\psi]\Phi^0+h.c.$ in eq.(\ref{5NJL}) 
has the form of the Yukawa interaction between
the bound state $\Phi^0$ and the free fermion scattering states.

Note the appearance of the 
color factors, $\sqrt{N_c}$ and $N_c$, in the second and third lines respectively. 
The third term, where from eq.(\ref{singlet}) we 
have $|\Tr\Phi|^2= N_c|\Phi^0|^2$, is the potential that creates the
semiclassical bound state with the $N_c$ enhancement.

The mass scale, $M$, is ultimately inherited by the bound state $\Phi$ from the coloron interaction.
The interaction introduces the scale $M_0$ and it will be
treated as a cut-off, as in the NJL model, hence $M < M_0$. 
The prefactor $M$ in eq.(\ref{Phi01})
is {\em a priori} arbitrary, but
we can swap $M$  for a dimensionless 
 parameter $\epsilon$ as:
\smallbbo
 M=\epsilon M_0
  \smallebo\normalsize
  $\epsilon $ should be viewed as part of the wave-function
of $\Phi$. In a variational
 calculation below,  (Section (\ref{varcalc})), we will see that $\epsilon =1$ 
 extremalizes the SKG effective Hamiltonian in a bound state (with negative $\mu^2$
 eigenvalue). Also,
 $\epsilon =0$ is the extremal value for the subcritical case.
 Hence, in the subcritical case $\Phi$ disappears and we are left
 with only unbound fermions.
 The reason is simple: $\epsilon$ rescales the coupling constant
 $\sim g_0^2\epsilon$ and the maximal value 
 of the coupling is therefore $\epsilon \rightarrow 1$.

\subsection{Fake Chiral Instability: The Need for the
Dynamical Internal Wave Function \label{instab}}

Consider the pointlike limit of eq.(\ref{5NJL}) 
and the semiclassical fields in eq.(\ref{4NJL}), 
where we replace $\Phi^0(x,y)\rightarrow \Phi^0(x)$,
and obtain,\footnote{ $\Phi(x)$ here {\em should not be confused with the factorized NJL interaction}, eq.(\ref{NJL2}), where $M_0^2$ is a right-sign non-tachyonic mass. 
Here $\Phi$ is not a pure auxiliary field, but rather
is physical. }
\smallbbo
\label{60NJL}
\backo
S'
\rightarrow \!\!\int_{x} \bl
\frac{g_0^2}{M_0^2}[\bar{\psi}_L\psi_{R}]_f [\bar{\psi}_R\psi_{L}]_{f}
+ \widetilde{M}^2 \Phi^0{}^\dagger\Phi^0\nonumbo \qquad\qquad
+ g^2_0\epsilon \sqrt{N_c}([\bar{\psi}_L\psi_{R}]_{f}\Phi^0 {+h.c.})
 \br
\smallebo\normalsize
where  $\widetilde{M}^2=g_0^2N_c\epsilon^2M_0^2$.

Eq.(\ref{60NJL})  
contains a wrong-sign (``tachyonic'') mass term,
implying a potential, $ -\widetilde{M}^2|\Phi|^2$.
This appears to generate spontaneous symmetry breaking
for any values of the underlying parameters $M_0$ and $g_0$
and the vacuum implodes.
A chiral vacuum instability is 
apparently an immediate, large effect of introducing eq.(\ref{act2})!

Such a conclusion is obviously
physically incorrect.  In
naively replacing $\Phi(x,y)$
with $\Phi(x)$ we have 
neglected the kinetic term of the internal wave-function, $|\partial_r\Phi|^2$
where  $r=(x-y)/2$.  This
opposes the instability like a repulsive interaction and will
stabilize the vacuum in weak coupling.  This is similar to the
stabilization of the classical Hydrogen atom by the Schr\"odinger wave-function.
A chiral instability can occur through competition of the repulsive internal wave-function
kinetic term and the attractive potential,
but will require a sufficiently large coupling, $g_0^2>g_{0c}^2$,
to drive it, and a quartic coupling to stabilize the vacuum.

We therefore must consider the internal dynamics of the non-pointlike bound state.
We note that this in the spirit of ref.\cite{hasenfratz}, but
 physically distinct, as the authors were arguing for a bare kinetic
term of the factorized NJL model of eq.(\ref{0NJL222}), while we are arguing for 
 a bare kinetic term of the bilocal field $\Phi(x,y)$, including the internal 
 coordinates $\sim x-y$.

\subsection{Bilocal Free Fields}

Consider a pair of massless particles of 4-momenta $p_1$ and $p_2$.  We have $p_1^2=p_2^2=0$.
and we can have two-body plane waves,
$\Phi(x,y) \sim \exp(ip_1x+ip_2y)$.  We  pass to the total
momentum $P=(p_1+p_2)$ and relative momentum  $Q=(p_1-p_2)$.
The plane waves become $\exp(iPX+iQr)$ where we define ``barycentric coordinates,'' 
\smallbbo
\label{bary}
X^\mu=\frac{x^\mu+y^\mu}{2},\qquad\;\; r^\mu=\frac{x^\mu-y^\mu}{2}, 
\nonumbo
\partial_x=\half(\partial_X+\partial_r),
\qquad\!\!\!\!\!
\partial_y=\half(\partial_X-\partial_r)
\smallebo\normalsize
Note that $P_\mu Q^\mu=p_1^2-p_2^2=0$. 
This implies that there is always a rest frame in which
$P=(P_0,0)$ and $Q=(0,\vvq)$. 
Hence, in the rest frame the dependence upon $\vec{X}$ and $r^0$ drops out.
If the particles are constituents of a bound state then this is the rest
frame of the composite particle.

To proceed we need  the generalized kinetic term of $\Phi(x,y)$
viewed as a bilocal field with an internal wave-function coordinate, $r^\mu$.
A free particle scattering state, $\Phi(x,y)$, 
composed of massless particles, will satisfy \cite{Yukawa},
\smallbbo
\partial^2_x\Phi(x,y) + \partial^2_y\Phi(x,y)=0
\smallebo\normalsize
or equivalently,
\smallbbo
\half \partial^2_X\Phi'(X,r) + \half\partial_r^2\Phi'(X,r) =0
\smallebo\normalsize
where $\Phi'(X,r)=\Phi(X-r,X+r)$.
 The bilocal field $\Phi(x,y)=\overline{\psi}_{R}(x)\psi_{L}(y)$
represents a ``bosonization'' of the pair of chiral fermions, as in chiral Lagrangians.
The equations of motion follow from the square of the free particle
Dirac equations, $(\slash{\partial})_{x}^2\psi_{R}(x)=0$ and $(\slash{\partial})_{y}^2\psi_{L}(y)=0$.
Note that we have chosen to describe the separation as $r$ which is the radius, where 
$ 2r=\rho\equiv (x-y) $ denotes
the separation of the particles. The choice of $r$ leads to more symmetrical
expressions in $r$ and $X$, and (somewhat) suppresses inconvenient factors of $2$.

The $\partial_x^2$ and $\partial_y^2$ are, in principle, independent for free fields.
We can also write an independent
Lorentz invariant equation,
\smallbbo
\backo\backo
\half\partial^2_x\Phi - \half \partial^2_y\Phi = 
\frac{\partial}{\partial X^\mu}\frac{\partial}{\partial r_\mu }\Phi'(X,r)=0 
\smallebo\normalsize
This can be viewed as a Lorentz invariant constraint
on the bilocal field. In Dirac constraint theory,
in a Hamiltonian formalism \cite{Dirac}, it is
the ``primary constraint.''
This constraint implies that
in the rest system  we have a stationary solution 
in which the ``relative time,'' $2r^0=\rho^0$, is removed, e.g.,
 \smallbbo
\label{factor000}
\Phi'(X,r)\rightarrow \Phi'( X^0,\vec{r}), \qquad \vec{P}=0,\;r^0=0.
 \smallebo\normalsize
 where $P_\mu = i\partial\Phi'/\partial X^\mu$ is the total 4-momentum of the state.
 This is ``stationary'' in the sense that the total 3-momentum
 of the system vanishes in the rest (barycentric) frame, $\vec{P}=0$, and
 we have a  wave-function, $\Phi'(t,\vvr)$
 where $t=X^0$, but depends only upon $\vvr$, with $r^0=0$.

 We can view the equations of motion as
arising from an action, subject to the constraint. 
With $M^2\Phi$ we choose the dimensionality of the field $\Phi$
to be that of a scalar field, i.e., $\sim $ (mass).  
We then write a dimensionless, free bilocal action,
\smallbbo 
\label{xxNJL0}
\backo
S_K=
\!\!
 \int_{xy}\!\!\! M^4 Z_0\bl |\partial_{x}\Phi |^2\!
+ \!|\partial_{y}\Phi |^2 \br
\smallebo\normalsize
The normalization factor $Z_0$ is defined below.
In the barycentric coordinates $\Phi(x,y) = \Phi'(X,r)$
and this becomes,
\smallbbo 
\label{xxNJL}
\backo
S_K=\half J
 \int_{Xr} Z_0M^4 \bl |\partial_{X}\Phi' |^2
+ |\partial_{r}\Phi' |^2 +...  \br
\smallebo\normalsize
where  $J=| \partial(x,y)/\partial(X,r)|=2^4$ is the Jacobian in passing from $(x^\mu,y^\nu)$
to the barycentric coordinates $(X^\mu,r^\nu)$.
(Our abbreviated notation for integrals is defined in Appendix \ref{SumNote}).

We can formally implement the constraint by adding
to the action a Lagrange multiplier, $\eta$,  while
preserving Lorentz invariance,
\smallbbo
\label{eqL}\backo\backo
S_\eta= \!\!\int_{Xr}\eta M^4 \left|\Tr
\frac{\partial\Phi'^\dagger{}}{\partial X^\mu}\frac{\partial\Phi'}{\partial r_\mu}\right|^2\;\;
\smallebo\normalsize
We will assume the Lagrange multiplier term is implicitly present in all the bilocal actions we 
write subsequently, and $\delta S_\eta/\delta \eta = 0$.
 
Following Yukawa \cite{Yukawa}  we consider,  in the barycentric coordinates, 
a factorization of  $\Phi'(X,r)$ 
as,
 \smallbbo
\label{af1000}
\sqrt{J/2}\; \Phi'(X,r)= \chi(X)\phi(r)
 \smallebo\normalsize
Obviously, an arbitrary function $\Phi(x,y)$ can be written in terms of
$\Phi'(X,r)$ since $\Phi(x,y)=\Phi(X+r,X-r)\equiv \Phi'(X,r)$. 
However an arbitrary  $\Phi(x,y)$
cannot generally be written in terms of a factorized pair as in eq.(\ref{af1000}).
An arbitrary $\Phi(x,y)$ can, however, be written as a sum over basis functions
of the form, $\sum\beta^{ij}\chi_i(X)\phi_j(r)$.

The pure factorization assumption of eq.(\ref{af1000}), however, can be viewed as
an ansatz
for the particular bilocal wave-function we expect for
composite particles. Then  $\chi(X)$ describes the center-of-mass
motion of the system and $\phi(r)$ describes the internal wave-function.

 The action eq.(\ref{xxNJL}) with the factorized field becomes,
 \smallbbo
\label{af2}
\backo \backo S=M^4
\!\!\int_{Xr} \!Z_0
\bl |\phi(r)|^2|\partial_X\chi(X)|^2
+|\chi(X)|^2|\partial_r\phi(r)|^2 \br
 \smallebo\normalsize
In the rest-frame the constraint implies,
\smallbbo
\frac{\partial}{\partial r^0}\Phi'(X,r)
=\frac{\partial}{\partial \vec{X}}\Phi'(X,r)=0
\smallebo\normalsize
 In the rest-frame we have the total 3-momentum, 
 $\vec{P}\sim \partial/\partial\vec{X}=0$, and
 time is carried by $X^0$. The internal wave-function  is independent of the relative time $r^0$ 
 and has
 dependence only upon $\vec{r}$.
 Hence we can define in this frame,  
 \smallbbo
\label{af20}
 \sqrt{J/2}\;\Phi'(X,r)=  \chi(X^0)\phi(\vvr).
 \smallebo\normalsize
 We can therefore integrate out $r^0$ and $\vec{X}$ (as defined in the rest frame),
 \smallbbo
 \int d^3\vec{X}= V_3\qquad \int dr^0 = T
 \smallebo\normalsize
We now choose 
 the normalization factor, $Z_0$, as,
 \smallbbo
 \label{Znorm}
 1= M\!\int \!\! dr^0 Z_0 = Z_0M T
 \smallebo\normalsize
 Here $Z_0$ is defined in the rest system with 
 a cut-off, $\int dr^0 \equiv T$.  
 
While this may seem like a frame dependent constraint,
we can in principle give a more detailed 
manifestly Lorentz invariant prescription for removal of the relative
 time integral in the kinetic term along the lines of Dirac's Hamiltonian constraint theory \cite{Dirac} (See \cite{bilocal} for the
 scalar field case that more directly leads to $Z_0$)

  These parameters are intuitively related to the physical properties of the state.
The spatial volume of the bound state is $\sim
 M^{-3}=\epsilon^{-3}M_0^{-3}$.  We view $M$ as part of the wave-function, and it is determined
 by extremalizing the Hamiltonian.  We find that, for supercritical
 coupling, $\epsilon \rightarrow 1$ and $M=M_0$, as seen in a simple
 variational calculation below.  For subcritical coupling the
 extremalization produces $\epsilon \rightarrow 0$, and indicates that $\phi(r)$ does not
 then involve the mass, $M_0$, but only momenta $\phi(\vec{q}\cdot\vec{r})$, and
 The  volume of $\phi(r)$ then  abruptly switches from $M_0^{-3}$ to $V_3$,
 from a compact bound state to an extended plane wave.
 $T$ in the bound state
 is essentially the transit time of the constituent pair in the core of the potential, and
 we therefore expect $T\sim M^{-1}_0$, and  $Z_0\sim 1$.

 With eq.(\ref{Znorm})  the action in the rest frame becomes,
 \smallbbo
  \label{af200}
\backo \!\! S_K\!=\!
{M^3}V_3\!\!\int\!\!dX^0 d^3r \bl\! |\phi(r)|^2|\partial_0\chi|^2
-|\chi|^2|\vec{\partial}_r\phi(r)|^2 \br
\smallebo\normalsize
Notice only the spatial derivative terms appear in the $\phi$
kinetic term.

To have the usual canonical normalization
of the $\chi$ field we therefore require the normalization of the dimensionless 
internal wave-function $\phi(\vec{r})$
 to be
 \smallbbo
 \label{phinorm}
 M^3\!\! \int d^3r\; |\phi(\vec{r})|^2 =1.
 \smallebo\normalsize
 Hence we are choosing the internal wave-function, factor field $\phi(r)$,
 to be dimensionless.

As a trivial example, consider an internal wave-function with $\phi(\vvr)=N\exp(2i\vvq\cdot\vvr)$.
 corresponding to a two body scattering state. The above normalization is then formally
 $M^3N^2\int d^3r=1$ and 
the action then becomes,
 \smallbbo
  \label{af3}
\backo \!\! S_K\!=\!
V_3\!\!\int\!\!dX^0 \bl |\partial_0\chi(X^0)|^2 - 4\vec{q}\;{}^2 |\chi(X^0)|^2\br
\smallebo\normalsize
 This is a state described by $\chi(X)$ 
 which 
 satisfies the equation of motion,
 \smallbbo
 \partial_0^2\chi + \mu^2\chi =0   \qquad \mu^2=4\vec{q}\;{}^2,
 \smallebo\normalsize
a zero 3-momentum two body
 scattering state of invariant mass $2|\vec{q}|=\mu $, with conventional normalization\footnote{The
 conventional free field normalization implies that the action reduces to the
classical massive particle form, $\int P_\mu dx^\mu=\int \mu \;dX^0$. }
 \smallbbo
 \chi = \frac{1}{\sqrt{2\mu V_3}}e^{i\mu X^0}  
 \smallebo\normalsize
 We discuss the bilocal scattering states further
in Appendix \ref{Free}.
A key point in this example is that $\vec{q}\;{}^2$, which appears as an ``invariant
mass,'' $-q^2=\mu^2$, is not really a mass at all, since the stress tensor trace
remains zero for massless particles. A true mass, such as associated with a bound state
of massless particles, 
requires a mass scale generated in the interaction 
and a nonzero stress tensor trace. In QCD this happens 
via the trace anomaly, which is $\propto \beta(g)/g$,
and similarly in Coleman-Weinberg dynamical symmetry breaking
\cite{CW}. See \cite{HillTA} for discussion of the role of the trace anomalies. In the present 
instance the mass scale $M_0$ could come indirectly
from a $\propto \beta(g_0)/g_0$ or other trace anomaly in the coloron dynamics.

 \subsection{Turning on the Interaction}

We now consider the interaction of eq.(\ref{5NJL}), particularly the last term.
 Here we have the full space-time dependence of 
 the propagator $D_F(x^\mu\!-\!y^\mu)=D_F(2r^\mu)$.
Hence the {\em quadratic action} of eq.(\ref{xxNJL}) for   $\Phi( X^0,\vvr) $  
in the rest frame becomes, 
\smallbbo 
\label{62NJL}
\backo\;
  JM^3 V_3 \!\! \int_{X^0\vvr}\bl \!  \half |\partial_{X^0}\Phi' |^2
-\half |\partial_{\vvr}\Phi' |^2 \nonumbo
+ JM^4 V_3 \int dX^0 d^3 r\; dr^0\; g_0^2N_cD(2r)|\Phi'|^2  \br
\smallebo\normalsize
(we consider the Yukawa and quartic interaction terms subsequently).

We again integrate over $r^0$ (the relative time), where
$\Phi'( X^\mu,\vec{r}) $ is constrained to have no dependence upon $r^0$.
However, now
in the third term we have,\footnote{
Note the argument of $D_F(2r^\mu)$ is the separation of the particles, $\rho^\mu=2r^\mu$
which leads to the factor of $\half$ in eq.(\ref{Ypot1}), which joins the overall factor of $\half$ in eq.(\ref{43}).  } 
\smallbbo
\label{Ypot1}
\int\!\! dr^0  D_F(2r)
= -\!\!\int\!\!  dr^0 \frac{d^4q}{(2\pi)^4}
\frac{1}{q^2-M_0^2}e^{i2 q_\mu r^\mu }
\nonumbo
= \half \!\!\int\!\!   \frac{d^3q}{(2\pi)^4}
\frac{1}{\vec{q}\;{}^2+M_0^2}e^{i2 q_\mu r^\mu }= -\half V_0(2|\vvr|)
\smallebo\normalsize
where the $\vec{q}$ momentum integral yields the familiar Yukawa potential, 
\smallbbo
\label{Ypot3}
 V_0(\rho) =-\frac{e^{-M_0 \rho}}{4\pi \rho}=-\frac{ e^{-2M_0 |\vvr|}}{8\pi |\vvr|}; \qquad \rho= 2|\vvr|
\smallebo\normalsize
In the limit of suppressing the
$\vec{q}\;{}^2$ in the denominator of eq.(\ref{Ypot1}) we obtain using $J=2^4$, 
and $\delta^3(\vec{r})=(4\pi r^2)^{-1}\delta(r)$:
\smallbbo
\label{Ypot2}
V_0(2r)\rightarrow -\frac{1}{M^2_0}\delta^3(2\vvr)=-\frac{1}{2\pi J M^2_0 r^2}\delta(r)
\smallebo\normalsize
From eq.(\ref{xxNJL}), including
color factors and introducing a quartic interaction (which is presently assumed to be 
$|\Phi|^4 \propto |\Phi(x,y)|^4$)
the resulting action in the barycentric frame is,
\smallbbo
\backo
\label{43}
S = \half J\epsilon^3M_0^{3} V_3\!\!
 \int\!\! dX^0 {d^3r}\bl  |\partial_{X^0}\Phi' |^2
- |\partial_{\vvr}\Phi' |^2 
\nonumbo \qquad \qquad
-  g_0^2 N_cM V_0(2r)|\Phi'|^2 -\frac{\lambda_0}{2}|\Phi'|^4 +... \br
\smallebo\normalsize
Note that since eq.(\ref{43}) is an action it contains $-(..)V_0(2r)$, while in eq.(\ref{Ypot3})
we see that $V_0=-(..)$ is defined with an intrinsic minus sign, hence this term is positive in the action and thus
attractive in a Hamiltonian.  Recall that with $\epsilon= 1$ we have $M=M_0$.

We emphasize the procedure of going to the rest frame is a ``gauge choice'' while the
theory remains overall Lorentz (``gauge'') invariant.
Note that $J\epsilon^3 M_0^{3}$
is a measure of the inverse volume of the bound state.
We can write the action (noting $M=\epsilon M_0$ and factors of $2$ that come from
the $\half J$ normalization of eq.(\ref{af1000}), and eq.(\ref{phinorm}))
in the format: 
\smallbbo
\label{51}
\backo\;\;\;
S = V_3\!\!
 \int\!\! dX^0 \bl |\partial_{X^0}\chi |^2 - |\chi |^2{\cal{M}}^2
- \frac{\tilde{\lambda}}{2}|\chi|^4\br
\nonumbo
\makebox{where,}\;\;\;
{\cal{M}}^2=M^3\int_{\vvr} \bl |\partial_{\vec{r}}\phi |^2+g_0^2N_c  M V_0(2r)|\phi|^2\br 
\nonumbo
=M^3\!\! \int_{\vvr} \bl |\partial_{\vec{r}}\phi |^2-g_0^2N_c M\frac{ e^{-2M_0 |\vvr|}}{8\pi |\vvr|}|\phi|^2\br 
\smallebo
We have assumed a simple form for the quartic term
where $\tilde\lambda$ absorbs an integral over $\phi(r)^4$ and factors of $J$.
The quartic interaction will be generated at loop level in Section \ref{loops}.

${\cal{M}}^2$ is the ``internal  Hamiltonian'' that describes the internal dynamics
of the bound state.  ${\cal{M}}^2$ appears in the action,
but contains no time derivatives since we have eliminated relative time. Therefore,
including a minus sign, ${\cal{M}}^2$ is a Hamiltonian for the static
internal wave-function $\phi(\vec{r})$. 

Using $V_0(2r)$ of eq.(\ref{Ypot1}) and extremalizing  ${\cal{M}}^2$ gives 
the SKG equation and it's eigenvalue,
 $\mu^2$:
  \smallbbo
\label{SKG}
\backo\!\!\!\!
-\bl\frac{\partial^2 }{\partial r^2}+\frac{2}{r}\frac{\partial }{\partial r}\br\phi(r)-g_0^2N_c M\frac{ e^{-2M_0 |\vvr|}}{8\pi |\vvr|}\phi(r) =\mu^2\phi(r)
 \smallebo\normalsize
 $\mu^2$ is then the physical mass  of the bound state, and the $\chi$ action in
 any frame is manifestly Lorentz invariant:
 \smallbbo
\backo\!\!\!
S = \!\!
 \int\!\! d^4X \bl |\partial_{X}\chi(X) |^2 - |\chi(X) |^2\mu^2
-\frac{\tilde{\lambda}}{2}|\chi(X)|^4\br.
 \smallebo\normalsize
 The Yukawa potential has a critical coupling, $g_0=g_{0c}$, 
 where the eigenvalue is then $\mu =0$.
 For $g_0>g_{0c}$ then $\mu^2<0$, and we have spontaneous symmetry breaking.
 
 A central feature of our formalism is that the internal coordinate, $\vec{r}$, 
 is always associated with the mass scale $M_0$ in the integrals,
 \smallbbo
 \sim M_0^3\int \! d^3r \;F(M_0\vec{r})
 \smallebo
with relevant factors of $\epsilon$ associated with $\phi$.
We can pass to a scale invariant internal coordinate variable, $u=M_0r$,
as in the case of the coloron model, as scale
invariant Hamiltonian $\hat{\cal{H}}$,
 \smallbbo
 \label{dim0}
\backo \!\!\! \hat{\cal{H}}=
{\cal{M}}^2/M_0^2
=
\epsilon^3 \!\! \int\!\! {d^3u}\bl |\partial_{u}\phi |^2 
-g_0^2N_c\epsilon \frac{e^{-2 u}}{8\pi u}|\phi|^2\br
\smallebo\normalsize
The eigenvalue of the associated SKG equation is then the dimensionless $\hat{\mu}^2=\mu^2/M_0^2$.
Postulating eq.(\ref{dim0}) as a general starting point,  we can argue that the mass
scale is entirely determined by the interaction, by passing back to 
$u=M_0r$ to match the Yukawa interaction.  In this way we make explicit
that it is the Yukawa interaction
that dictates the mass scale, $M_0$.

 \subsection{The Induced Bound State Yukawa Interaction}
 
The Yukawa interaction of the bound state with
the free scattering state fermions is now induced from the second term, $S_Y'$,  in eq.(\ref{5NJL}). We have,
noting  eq.(\ref{af20}):
\smallbbo 
\label{5NJL000}
\backo
S_Y' =
g^2_0\sqrt{N_c}M^2\!\!\int_{xy}\!\![\bar{\psi}_L(x)\psi_{R}(y)]_{f}  D_F(x-y)\;\Phi^0(x,y) {+h.c.}
\nonumbo
=  \sqrt{2N_cJ} g^2_0M_0^2\times
\nonumbo\;\;
\int_{Xr}\!\![\bar{\psi}_L(X\!+\!r)\psi_{R}(X\!-\!r)]_{f}D_F(2r)\;\chi(X)\phi(\vvr) {+h.c.}
\smallebo\normalsize

Consider the pointlike limit of the potential,
eq.(\ref{5NJL000})
\smallbbo
\backo
S_Y'
\rightarrow
\nonumbo
\backo
 g^2_0\sqrt{N_c}M^2\!\int_{xy}\![\bar{\psi}_L(x)\psi_{R}(y)]_{f} \frac{\delta^4(x-y)}{M_0^2}\;\Phi^0(x,y) {+h.c.}
 \nonumbo
 \rightarrow
 g^2_0\epsilon^2\sqrt{N_c}\!\int_{x}\![\bar{\psi}_L(x)\psi_{R}(x)]_{f} \;\Phi^0(x,x) +h.c.
 \nonumbo
  \rightarrow
 g^2_0\epsilon^2\sqrt{2N_c/J}\!\int_{x}\![\bar{\psi}_L(x)\psi_{R}(x)]_{f} \;\chi(x) \phi(0) +h.c.
\smallebo\normalsize

We therefore see that the induced Yukawa coupling in the pointlike limit
to the field $\chi(x)$ is:
\smallbbo
\label{gy}
g_Y= \hat{g}_Y\phi(0)\qquad  \hat{g}_Y\equiv g_0^2\epsilon^2\sqrt{2N_c/J} 
\smallebo\normalsize

This is a significant result and fundamentally different than the pointlike
NJL result.  We have taken the pointlike limit of the potential as in the
NJL model, but obtain a result that is dependent
crucially upon the internal wave-function $\propto \phi(0)$. The implication
 is that a strong coupling, $g_0^2$,  can produce, in principle, a small
Yukawa coupling if $\phi(0)<\!\!< 1$. 
In the usual pointlike NJL model the induced Yukawa coupling runs  to smaller
values in the IR, but it does so only logarithmically via the RG.  Here the behavior
of $\phi(0)$ is a suppression of $g_Y$  that is power-law near the critical coupling.

Alternatively,  in terms of an extended potential, $V_0(2r)$,
\smallbbo
\label{extyuk}
\backo
S'
\rightarrow
-\sqrt{N_cJ/2} g^2_0M^2\times
\nonumbo\;\;
\!\int_{X\vvr}\!\![\bar{\psi}_L(X^+)\psi_{R}(X^-)]_{f}\chi(X)V(2r)\phi(\vvr) {+h.c.}
\smallebo\normalsize
where $X^{\mu\pm}=(X^0, \vvX \pm \vvr)$
and we have the induced Yukawa coupling 
\smallbbo
g_Y =-g_0^2\epsilon^2 \sqrt{N_cJ/2}\; M^2_0\int\!\! 4\pi r^2 dr  V_0(2r)\phi(r)\!\!
\smallebo\normalsize
We discuss some further issues of the induced Yukawa interaction
in Appendix \ref{yukint}.

\section{The Schr\"odinger-Klein-Gordon (SKG) Equation}

While formally similar to the non-relativisitic Schr\"odinger equation, the SKG
equation has key physical differences:
(1) the potential has dimension (mass)$^2$, rather than energy;
(2) the eigenvalue describes resonances for positive $\mu^2$; 
(3)  ``tachyons'' occur  for negative $\mu^2$, which implies  vacuum instability
and spontaneous symmetry breaking. 
Mainly the SKG Hamiltonian is amenable to variational calculations as we show below. 
We presently give some examples of solutions and stress
some subtleties. Much can be done to refine and extend this discussion.
The solutions allow the computation of the induced Yukawa coupling
of the bound state to free fermions, $g_Y$, via the wave-function
at the origin, $\phi(0)$.

 \normalsize
 A negative eigenvalue of the Schr\"odinger  equation defines our conventional
 view of a nonrelativistc bound state, however, in the relativistic case 
 for a pair of chiral fermions the SKG equation implies a negative $\mu^2$.   
 This is, of course, the
 behavior of $\Sigma$-model in QCD and the BEH boson in the standard model
 and requires
 additional physics to stabilize the vacuum, such as quartic interactions. 
 Hence, the general result is that 
 a scalar bound state of massless
 chiral fermions in the symmetric (unbroken) phase must 
 either be an unstable resonance (subcritical coupling and positive $\mu^2$), which decays rapidly to its
 constituents,
 or tachyonic (supercritical coupling, negative $\mu^2$) leading to a chiral instability of the vacuum.
 
 From the action of eq.(\ref{51})
 the  ``internal Hamiltonian'' ${\cal{M}}^2 $ describes the $\phi(r)$ 
 field, and since there is no time derivative for $\phi(r)$,
${\cal{M}}^2 $ is just $-1$ times the $\phi$ action.  
The most negative eigenvalue occurs when $g_0^2\epsilon$ is maximal, $\epsilon=1$
as seen below from a variational calculation.

\subsection{Exact Criticality of the Yukawa Potential}

The coloron model  furnishes a  direct UV
completion of the NJL model.  It leads to an SKG potential of the Yukawa form
which has a critical coupling, $g^2_0=g^2_{0c}$.
 The critical coupling is the nonzero value of $g_0^2$
for which the eigenvalue $\mu^2$ is zero.
We wish to determine $g_{0c}^2$. 

The criticality of the Yukawa potential in the nonrelativistic
Schr\"odinger equation is widely discussed in the literature in the context of ``screening''
(see \cite{Edwards} and references therein).
The nonrelativistic Schr\"odinger equation $r=|\vvr|$ is:
 \smbbo
 \label{SKG01}
-\nabla^2\psi - 2m_e\alpha\frac{e^{-\mu r}}{r}\psi=2m_eE
 \smebo
 with $m_e$ the electron mass, 
and eigenvalue $E=0$ occurs for a critical screening
with $\mu=\mu_c$. A numerical analysis yields,
\cite{Edwards},
 \smbbo
 \label{Edwards2}
\mu_c= 1.19061 \;\alpha m_e.
 \smebo
For the spherical SKG equation in the coloron model eq.(\ref{SKG}) we have from the Hamiltonian,
  \smallbbo
\label{SKG00}
\backo
-\nabla^2\phi(r)-g_0^2N_c M_0\frac{ e^{-2M_0 |\vvr|}}{8\pi |\vvr|}\phi(r) =\mu^2\phi(r)
 \smallebo\normalsize
where we assume  $\epsilon=1$ as determined by a variational calculation below.
 
 We can obtain the critical coloron model coupling constant by 
comparing, eq.(\ref{SKG01}) and eq.(\ref{SKG00}). We have,
\smallbbo
2m_e\alpha \rightarrow  g_0^2 N_c M_0/8\pi
\nonumbo
\mu_c\rightarrow 2M_0
\smallebo
substituting into eq.(\ref{Edwards2}),  
$ 2M_0= 1.19061 (g_0^2 N_c M_0/16\pi)$ and therefore,
\smallbbo
\label{exact}
\left.\frac{g_0^2N_c}{8\pi^2}\right|_c= \frac{4}{(1.19061)\pi} =1.06940
 \smebo
By comparison, the loop level NJL critical value of eq.(\ref{NJL3}) is, 
\smallbbo
\label{critc2}
\left.\frac{g_0^2N_c}{8\pi^2}\right|_{NJLc}=1.00
\smallebo
Hence, we see that the NJL quantum critical coupling has a remarkably
similar numerical value to the classical critical coupling.
(It is beyond the scope of the present paper to understand why these
are not identically equal!) 

Note that in what follows we will use a different definition
of the coupling,
\smbbo
\kappa =\frac{g_0^2N_c}{4\pi}\qquad \kappa_c= 2\pi 
\smebo
where we quote the implied NJL critical value $\kappa_c$.

\subsection{Rectangular Potential Well}

Presently, we consider a generic potential, $W(r)$, to write,
 \smallbbo
 \label{hamone}
\backo\;\;\;
{\cal{M}}^2=M_0^3 \int_{\vec{r}} \bl (\partial_r\phi(r))^2 +g_0^2N_c W(r)|\phi|^2\br  
\smallebo\normalsize
from which we obtain the SKG equation for a spherically symmetric wave-function 
with eigenvalue $\mu^2$: 
 \smallbbo
\backo
 -\nabla^2 \phi(r) +g_0^2N_c W(r)\phi =\mu^2\phi(r)  
\smallebo\normalsize
where $\nabla^2 = \partial_r^2 -(2/r)\partial_r$.
As a warm-up exercise we turn to  
the rectangular potential well,
 \smallbbo
 W(r) = -M^2_0\theta(R_0 -r)
 \smallebo\normalsize
  With $\phi(r)=u(r)/r$, the SKG equation becomes (here $\epsilon = 1$),
   \smallbbo
\label{SKG10}
\backo
\backo
-\frac{\partial^2 }{\partial r^2}u(r)-g_0^2N_c M^2_0 \theta(R_0 -r)u(r) =\mu^2u(r).
 \smallebo\normalsize

 For super-critical coupling, $g\gta g_c$, we have 
 a solution that is finite at $r=0$, and exponentially attenuating at large $r$ (hence
 normalizable),
 \smallbbo\label{anztz}
 \backo\!\!\!
 u(r) = A\sin(kr)\theta(R_0\!-\!r)+ Be^{-|\mu|(r\!-\!R_0)}\theta(r\!-\!R_0)
 \smallebo\normalsize
 where the eigenvalue is determined in the well as,
 \smallbbo
 \label{73}
\mu^2 =k^2-g_0^2N_cM_0^2
 \smallebo\normalsize
 The matching boundary conditions of the field value and derivative at $r=R_0$ 
 imply,
 \smallbbo\label{74}
 k^2\cot^2(kR_0)=k^2-g_0^2N_cM_0^2
 \nonumbo
  \sin^2(kR_0)=\frac{k^2}{k^2+ |\mu|^2}
 \smallebo\normalsize
 
 With critical (supercritical) binding we have $\mu^2=0$ ($\mu^2<0$).
 The critical value of $g=g_c$  therefore corresponds to,
 $k=\pi/2R_0$,  and the critical coupling implies,
 \smallbbo
 \label{75}
 \frac{\kappa_c}{2\pi}\equiv \left.\frac{g^2_{0c}N_c}{8\pi^2 }\right|_c = \frac{1}{32 M^2_0R^2_0} 
 \smallebo\normalsize
 We can use the square well as an approximation
 to the Yukawa potential.  In order for this to reproduce the critical value
 of the NJL model, close to the classical Yukawa potential result, 
 we require,
 \smbbo
 \label{apprx}
 R_0= \frac{1}{4\sqrt{2} }\;M_0^{-1}
 \smebo
 This is a narrow rectangular potential approximation as we would expect from the
 $e^{-2M_0}/r$ Yukawa form. 
 We remark that a ``frying pan'' potential, in which $R_0> \!\! >M_0^{-1}$, 
 can have an arbitrarily small critical coupling. However, typical gauge boson exchange
 potentials will have $R_0 \sim  1/M_0$ as indicated by this result.

  For the near critical coupling case $|\mu| <\!\!< 1 $ we see from eq.(\ref{74}) that 
  $\sin(kR_0)\rightarrow 1$, hence from eq.(\ref{anztz}),
  ${B}\rightarrow {A}$,  and therefore,
 \smallbbo
 \backo\!\!\!\!\!\!
 A^{-1}\phi(r) \! \approx \! \frac{1}{r}\theta(R_0\!-\!r)\sin\left(\!\frac{\pi r}{2R_0}\!\right)\!
 + \!\frac{e^{-|\mu|(r-R_0)}}{r}\theta(r\!-\!R_0)
 \smallebo\normalsize
 where $A$ is the normalization of $\phi(r)$ and $\mu\approx 0$. 
 We see that the critical solution $\mu= 0$ has a tail for large $r$, $\phi(r)\sim 1/r$.
 We include a small nonzero  $|\mu|$ in the last term 
 as an IR cut-off on the normalization integral. We can approximate the coupling
 by the critical value in the near critical case.
 
 Recall that the normalization is defined by eq.(\ref{phinorm}), 
 \smallbbo
 \backo\!\!\!
 1= M_0^3 \int_0^\infty \!\!\! 4\pi r^2 |\phi(r)|^2 dr 
 \approx 2\pi A^{2}  M_0^3\bl  R_0 + |\mu|^{-1}\!\br
 \smallebo\normalsize
 We see that near criticality the normalization is dominated by the
 large distance tail, 
 \smallbbo
 A=\frac{ M_0^{-3/2}|\mu|^{1/2}}{\sqrt{2\pi}}
 \smallebo\normalsize
 The result for $\phi(0)$ is then 
 \smallbbo
 \backo\!\!\!
 \phi(0)= \lim_{r\to 0}\frac{A}{r}\sin\left(\!\frac{\pi r}{2R_0}\right)=\frac{\sqrt{\pi} |\mu|^{1/2}}{\sqrt{8}M_0^{3/2} R_0} 
 =\bl\frac{4\pi |\mu|}{M_0 }\br^{1/2}
 \smallebo\normalsize
 where in the last term we inserted the NJL critical value $M_0R_0= 1/4\sqrt{2}$.

 The key observation is that $\phi(0)\sim (|\mu|/M_0)^{1/2}$ becomes small 
 as we approach the critical value where $|\mu| <\!\!< M_0$.
 This has the effect of suppressing the induced Yukawa coupling by a factor
 of $\phi(0)$ from eq.(\ref{gy}) (for $\epsilon = 1$):
 \smallbbo
\label{gysquarewell}
\backo
g_Y= g_0^2\sqrt{N_c/8}\;\phi(0)
\smallebo\normalsize
In contrast,
the renormalized Yukawa coupling  the NJL model diverges as a Landau pole
at the scale $M_0$, and then evolves logarithmically, as in eq.(\ref{NJL30}). 
Subsequently it can be matched onto the
full RG evolution \cite{BHL}, which leads to the IR fixed point \cite{PR}, hence in NJL
model the RG evolution is slow.   
In the present semiclassical case the logarithmic evolution is replaced by the 
more rapid power-law evolution 
$\phi(0)\propto \sqrt{|\mu|/M_0}$.

It is instructive to derive the square well potential results in an
approximation that we will develop in
the next section on variational methods.  We can set $R_0 = 1$
and we then assume $R_0M_0 =M_0=1/(4\sqrt{2}) \equiv f  $ from eq.(\ref{apprx}) 
for our approximation to the Yukawa potential.
We write an approximate trial wave-function for $\phi(r)$,
as:
\smallbbo
\backo
\phi(r)= \frac{u(r)}{r}
\nonumbo
\backo
u(r)= (r+Br^3)\theta(1-r)+Ce^{-\mu(r-1)}\theta(r-1)
\smallebo
Demanding the differentiability of $u(r)$  at $r=R_0$
(i.e., $u(r)$ is a C$^1$ function)
we have,
\smallbbo
B=-\frac{1+\mu }{3+\mu }\qquad C= \frac{2}{3+\mu } 
\smallebo
and the Hamiltonian has dimension $1/R_0^2$, and  in the field $u(r)$ is
 \smallbbo
\backo\!\!\!\!
{\cal{M}}^2= 4\pi A^2 {f^3} \int_0^\infty\!\!\! dr \bl |\partial_r u(r)|^2 +g_0^2N_c W(r)|u(r)|^2\br  
\nonumbo
=
4\pi A^2 f^3 \bl \frac{8}{15} + \frac{2}{45}\mu 
-g^2 N f^2\bl \frac{68}{315} -\frac{64}{945}\mu \br  \br
\smallebo
where $ W(r) = -f^2\theta (1-r)$
(note, the formal difference with the ${\cal{M}}^2(\phi(r))$ expression 
of eq.(\ref{hamone})
is a total derivative that integrates to zero).
The normalization is, 
\smallbbo
\backo\backo
{A}^2= \bl\! 4\pi f^3 \int_0^\infty\!\!\! dr |u(r)|^2\!\br^{-1}\!\!
=\frac{9\mu}{8\pi f^3}-\frac{12\mu^2}{35f^3\pi}+ {\cal{O}}(\mu^3)
\smallebo

The normalized expectation value of the Hamiltonian is obtained, restoring factors of
$R^{-1}_0=M_0/f$,
and  $\mu \rightarrow  \mu R_0=f\mu/M_0$,
\smallbbo
\backo
{\cal{M}}^2 = 4\pi A^2f \bl \frac{8}{15}M_0^2 + \frac{2}{45}f\mu M_0
\nonumbo
\qquad
-g^2 N f^2\bl \frac{68}{315}M_0^2 -\frac{64}{945} f \mu M_0 \br  \br
\nonumbo
=\mu M_0\bl \frac{48\sqrt{2}}{5}+\frac{\mu}{5M_0}
-  \frac{17\sqrt{2}}{35}\pi\kappa +\frac{4\mu}{105M_0}\pi\kappa \br
\smallebo
where $\kappa=g^2N/4\pi$.  Note the suppression of  ${\cal{M}}^2$,
which would be $\sim M_0^2$ without the tail of the wave-function, but
has now become $\sim \mu M_0$ due to the infrared dilution.

The critical coupling corresponds to $\mu=0$ and implies
\smallbbo
\kappa=\frac{336}{17\pi}  = 6.2913, \qquad
g^2N/8\pi^2=1.0013,
\smallebo
in excellent agreement with the NJL result (which was an input of $1.000$
to define $f$; this calculation essentially amounts to replacing the
$\sin(kr)$ by the first two terms in its series expansion).

Another  consideration is the fine-tuning for a hierarchy, $\mu/M_0 <\!< 1$. 
The result for ${\cal{M}}^2$
is the eigenvalue $-|\mu|^2$.  Therefore we have, keeping
the leading terms in $\mu^2$ in the Hamiltonian,
\smallbbo
-\mu^2
\approx \mu M_0\bl \frac{48\sqrt{2}}{5}
-  \frac{17\sqrt{2}}{35}\pi\kappa  \br 
\smallebo
hence the linear relation:
\smallbbo
\kappa =   6.2913 +0.90499\frac{\mu}{M_0}
\smallebo
This means that fine-tuning to achieve  a small
value of $\mu/M_0$ is highly suppressed, $\delta \kappa/\kappa \sim \mu/M_0$
as opposed to $
\mu^2/M_0^2$ without the dilution. This is due to the
approximate the linearity
of this relationship.

\subsection{Variational Calculations \label{varcalc} }

 A solution to the SKG equation for the eigenvalue
 can be computed approximately in a variational calculation.  
  In using variational methods it is important that the ansatz for the field configuration
  be a continuous function of the field value and its first derivative (a C$^1$ function).
  Discontinuities in the kinetic term lead to unwanted spikes in the Hamiltonian and affect
  the energetics. It is also important, if possible, to use known properties of the solution for the asymptotics. 
  
  We demonstrate this with a crude approximation in the present section that shows
  $\epsilon=1$ is the extremal solution for a bound state. We then illustrate a
   refined ``spline'' approximation that obtains precise (and interesting) results in 
   Section \ref{splinea}.
  
 First we recall $M=\epsilon M_0$, as the mass scale in the ansatz, 
reintroducing the parameter $\epsilon$.
As in the NJL model, we view this as a UV cut-off theory where
the largest mass scale is the coloron mass $M_0$, 
the largest mass scale at which the static potential approximation is applicable.
Hence we require $\epsilon \leq 1$.
$\epsilon $ is seen to multiply the underlying coupling constant, $\tilde{g}_0^2 = \epsilon g_0^2$.
The largest value of the ``effective coupling,''  $\tilde{g}_0^2$,
is therefore $g_0^2$, hence $\epsilon=1$ implies the smallest
possible critical value of the underlying coupling $g_0^2$.  
However, we can view $\epsilon$ as part of the wave-function ansatz, and allow the variational
calculation of the bound state mass to produce $\epsilon=1$ to minimize the Hamiltonian,
${\cal{M}}^2$, of eq.(\ref{51}) with the Yukawa potential of eq.(\ref{Ypot3}). 
  
  We consider a simple example of assuming
 an ansatz that is a Hydrogenic 
 wave-function, $\widetilde{\phi}(r) = Ae^{-Mr}$, with $M=\epsilon M_0$ and $\epsilon$ as the variational
 parameter, and $M_0$ is the scale of the Yukawa potential.  
 This cannot be an accurate description near criticality where the eigenvalue $\mu^2$ is small because
 it lacks the large distance tail  $\propto e^{-\mu r}/r$ for small $\mu$ (we'll include that subsequently).
 However, this gives a  rough approximation and establishes $\epsilon\rightarrow 1$ dynamically.

 The normalization of the ansatz is defined in  eq.(\ref{phinorm}),
  \smallbbo
  \backo
 1= 4\pi A^2\epsilon^3 M_0^3\int_0^{\infty}\!\!\!\! e^{-2\epsilon M_0r}r^2 dr   \qquad A^2=\frac{1}{\pi }
 \smallebo\normalsize
The Hamiltonian ${\cal{M}}^2$ of eq.(\ref{51}) with $M=\epsilon M_0$ is 
 \smallbbo
\label{510}
{\cal{M}}^2=\epsilon^3M_0^3\int_{\vvr} \bl |\partial_{\vec{r}}\phi |^2+g_0^2N_c \epsilon M_0 V_0(2r)|\phi|^2\br 
\smallebo\normalsize
where 
 $V_0(2r)=- e^{-2M_0 r}/{8\pi r}$
as in eq.(\ref{Ypot3}), with the fixed coloron mass $M_0$ (no $\epsilon$ factor is present in $V_0(2r)$!).
 We then compute the eigenvalue $\mu^2 = {\cal{M}}^2$ as a function
 of $\epsilon$ and $\kappa$:
 \smallbbo
 \label{epsvar}
 {\cal{M}}^2=A^2  \epsilon^3 M_0^3
 \int_{\vvr} \bl |\partial_{\vec{r}}\phi |^2-\frac{\kappa\epsilon M_0}{2} \frac{e^{-2M_0r}}{r} |\phi|^2\br 
 \nonumbo
 =4 \epsilon^3 M_0^3
 \int_0^\infty\bl \epsilon^2 M_0^2 e^{-2\epsilon M_0 r}-\frac{\kappa\epsilon  M_0}{2} \frac{e^{-2(1+\epsilon) M_0r}}{r}\br r^2 dr
\nonumbo
=M_0^2\bl \epsilon^2 - \frac{\kappa \epsilon^4 }{2(1+\epsilon)^2}\br;\;\;\;   \qquad \kappa = \frac{N_cg_0^2}{4\pi}
\smallebo\normalsize

{
\begin{figure}
	\centering
	\includegraphics[width=0.48\textwidth]{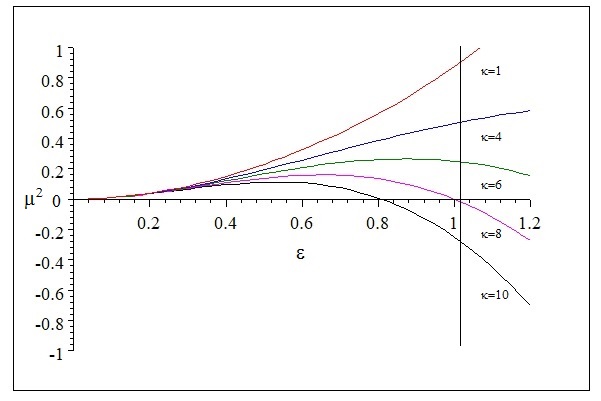}
	\vspace{-0.0in}
	\caption{${\cal{M}}^2=\mu^2$  of eq.(\ref{510}) is plotted vs. $\epsilon$, with $M_0=1$
	for values of $\kappa= (1,4,6,8,10)$. 
	The critical  coupling is the value of $\kappa$ for which a massless bound state 
occurs, ${\cal{M}}^2=0$, hence where the $\kappa=8$ curve intersects with 
the $\epsilon =1$ vertical line. For supercritical $\kappa>\kappa_c$ we have $\epsilon= 1$ 
and the eigenvalues $\mu^2$ are negative; For subcritical $\kappa<\kappa_c$ we have $\epsilon= 0$ 
and the eigenvalues $\mu^2$ are positive.  Resonances can exist in scattering states 
with eigenvalues $\vec{q}\;{}^2$
as in eq.(\ref{af3}) with plane wave volume normalization.}
	\label{fig1}
\end{figure}
}

In Fig.(\ref{fig1}) we plot a family of curves of ${\cal{M}}^2=\mu^2$ for various values of  $\kappa$ 
as function of $\epsilon$.
We see  that the extremal (smallest) value for positive $\mu^2$ is
$\mu^2=0$ and occurs for any $\kappa< 8$. On the other hand,
for $\kappa> 8$ the extremal
(most negative) value of $\mu^2={\cal{M}}^2 $
occurs when $\epsilon\rightarrow 1$.  Hence, the critical coupling is $\kappa=8$ and we have
 \smallbbo  \frac{\kappa}{2\pi} =\frac{ g^2_{0c}N_c}{8\pi^2}= 1.27,
 \smallebo 
 compared to the NJL critical value $1.00$.
 
A negative eigenvalue  $\mu^2 ={\cal{M}}^2$ is determined 
 where the corresponding $\kappa>8$ curve intersects the vertical line at $\epsilon =1$,
 that is,
\smallbbo
 \label{eigenvar}
\mu^2
=M_0^2\bl 1 - \frac{\kappa }{8}\br.
\smallebo\normalsize
Since for subcritical, $\kappa <8$, the eigenvalue $\mu^2=0$,
we see that is a second order phase transition behavior in $|\mu|$ vs $\kappa$.
 
The variational ansatz isn't too far off, however,
this gives a false value for the  normalized trial wave-function 
 at the origin for critical coupling,
 \smallbbo
 \phi(0)=  \frac{1}{\sqrt{\pi} } 
 \smallebo\normalsize
The reason is, of course, that this variational calculation truncates the $1/r$ tail at large $r$,
which significantly affects the normalized $\phi(0)$.
We now do a refined calculation that demonstrates the effects of the large distance
tail of $\phi(r)$.

\subsection{Spline Approximation \label{splinea}}

We can improve the method by constructing an ansatz for $\phi(r)$ that 
implements the short distance $e^{-M_0r}$, but
approximates the correct
asymptotic form, $\sim e^{-\mu r}/{r}$,
at large $r$.  

For $\mu$ small (near critical coupling) this is approximately $1/r$, and we then have
a convenient  ``spline'' (power + exponential) for $\phi(r)$, 
\smallbbo 
\backo
\label{ansatz1}
\phi(r) = A\left( e^{-M_0r}\theta(1-M_0r)+ \frac{e^{-1}}{M_0r} \theta(M_0r-1) \right)
\smallebo\normalsize
where the step function is $\theta(x) =1\;(=0)$ for $x>0\;(<0)$.
The spline is differentiable ($C^1$) which avoids the discontinuity in the value of the
function, as well as $\partial\phi \sim \delta(r-M)$ in its first derivative, which
would lead to ``energy spikes'' in the kinetic term.

With this definition of $\phi(r)$, however, we have an infrared divergent normalization
and kinetic term integrals
which require a cut-off with a small mass $|\mu|$.
The cut-off is equivalent to redefining the 
second term in eq.(\ref{ansatz1}) as $\rightarrow (e^{-1}/M_0 r) \theta(M_0  r-1)\theta(1-|\mu|r)$,
which allows  the integrals to run from $0$ to $\infty$. 
The cut-off at $|\mu|^{-1}$ imitates the desired true asymptotic form, $\sim e^{-|\mu| r}/r$,
and maintaining the continuity $\phi(r)$
at $r=M_0^{-1}$.
 
 Now, however, we have introduced a discontinuity at $r=|\mu|^{-1}$ and an unwanted $\delta(r-|\mu|^{-1})$
 in the kinetic term.  To remedy this we can simply extend the spline with a
 pure decaying exponential, as:
 \smallbbo
\label{ansatz2}\backo
\phi(r) = A\bl e^{-M_0r}\theta(1-M_0r)+ \frac{e^{-1}}{M_0r} \theta(M_0r-1)\theta(1-|\mu| r)
\nonumbo \qquad + \frac{|\mu|}{M}e^{-|\mu| r}\theta(|\mu| r-1) \br
\smallebo\normalsize
This is now a  differentiable function with value and derivatives matching at $Mr=1$
an $|{\mu}| r =1$. and we have eliminated the  IR divergence and can use this
for any value of $|\mu|$.

We have experimented with several ansatze and prefer the simplicity of eq.(\ref{ansatz2}). We find splining to an asymptotic function $\sim e^{-|\mu| r}/r$ 
leads to more cumbersome integrals.
Note the discontinuity at the origin,  $\partial_r \theta(r) e^{-M_0r}\sim \delta(r)$,
receives a factor of zero in the integrals due to the $r^2 dr$, hence causes no
discontinuity problem at $r=0$.
One might worry that this wave-function is not $C^2$ and cannot evidently satisfy the
equation of motion, however this is a variational ansatz and none of it satisfies the
equation of motion, yet it can produce reasonably reliable Hamiltonian expectation values.

The normalization of the trial wave-function 
of eq.(\ref{ansatz2}) must satisfy eq.(\ref{phinorm}), defined by $A$ given by,
\smallbbo
  \backo\!\!\!\!
 1= 4\pi A^2M_0^3\bl \int_0^{M_0^{-1}} \!\!\!\!e^{-2M_0r}r^2 dr +\int_{M_0^-1}^{|\mu|^{-1}}\!\!\!\!
 \left(\frac{e^{-1}}{Mr}\right)^2 r^2 dr  
 \nonumbo +
 \int_{|\mu|^{-1}}^\infty \left(\frac{|\mu|}{M}\right)^2 e^{-2|\mu| r} r^2 dr\!\br    
\smallebo\normalsize
hence, defining $x=|\mu|/M_0$ and $M_0=1$ we have 
\smallbbo
\label{A}
  \backo\!\!\!\!
 A= \frac{1}{\sqrt{\pi}}\bl 1+9e^{-2}\left(\frac{1}{x}-1\right)\br^{-1/2}   
\smallebo\normalsize
We see this is dominated by the tail of the wave-function for small $|\mu|$ as:
\smallbbo
\label{tail}
  A\approx \frac{e}{3\sqrt{\pi}}\bl\frac{|\mu|}{  M_0 }\br^{1/2}
\smallebo
 We then compute ${\cal{M}}^2$ as (recall $\kappa = {g_0^2N_c}/{4\pi}  $):
 \smbbo
 {\cal{M}}^2=4\pi  M_0^3\bl
 \int_{\vvr} \bl |\partial_{\vec{r}}\phi |^2-\frac{\kappa M_0}{2} \frac{e^{-2M_0r}}{r} |\phi|^2\br 
 \nonumbo
 =
 4\pi A^2M_0^3\bl \int_0^{M_0^{-1}} \!\!\!\!\bl M_0^2 e^{-2M_0r} -\frac{\kappa M_0}{2} \frac{e^{-4M_0r}}{r}\br r^2 dr
 \nonumbo 
 +  \int_{M_0^{-1}}^{|\mu|^{-1}}\!\! 
 \bl \frac{e^{-1}}{Mr} \br^2 \bl \frac{1}{r^2} 
 -\frac{\kappa M_0}{2} \frac{e^{-2M_0r}}{r}\br
 r^2 dr \br  
 \nonumbo 
 +  \int_{|\mu|^{-1}}^\infty\!
 \bl \frac{|\mu|}{M}e^{-\mu r} \br^2 \bl \mu^2
 -\frac{\kappa M_0e^{-2M_0r}}{2r}\br\br
 r^2 dr \br
 \smallebo
 We obtain 
 the result:
 \smallbbo 
 \label{M}
 \backo
  {\cal{M}}^2 =
\pi A^2 \bl1-e^{-2}(1-x)
 -\kappa\bl 2e^{-2}(\makebox{Ei}(1,2)-\makebox{Ei}(1,2/x))
 \nonumbo
 \qquad
+\frac{1}{8}\left(1-e^{-4}\right)
+\half\frac{3x^2+2x}{(1+x)^2}e^{-2(1+x)/x}\br \br
\smallebo
 where the exponential integral is $\makebox{Ei}(x)= \int_{-\infty}^x (e^t/t)dt$.

We proceed to compute ${\cal{M}}^2(\kappa)$.
We are interested only in negative eigenvalues for the compact wave-function. 
We input a trial value of $|\mu|/M_0$ (where we set $M_0=1$) 
and compute ${\cal{M}}^2=\mu^2$ for given values of $\kappa$. 
This leads to the family of curves seen in Fig.(\ref{figvarfamily}).

The eigenvalue  $\mu^2=-|\mu|^2={\cal{M}}^2$ is
produced by this formula. However, we have used as input $|\mu|/M_0$
to define the large distance tail of the wave-function.
The resulting output $\mu^2=-|\mu|^2$ {\em must self-consistently match the input value} $|\mu|$.
 Fig.(3) shows ${\cal{M}}^2$ with $M_0=1$ plotted for given values of $\kappa$ 
 vs the input $|\mu|$ with $M_0=1$.  
Self-consistent solutions occur where the given $\kappa$ curve
intersects a $\mu^2$ curve (thick curve Fig.(3)).  
The intersections implicitly determine $\kappa$ for any consistent
value of input $|\mu|$.  

The logic of the plots is that we are seeking 
the value of $\kappa$ that best yields a self-consistent solution for given $\mu$ (as 
in the critical coupling determination where we sought the best value of 
$\kappa$ that yields $\mu=0$).
This then gives us the value of $\mu$ for a given input value of $\kappa$.
In Fig.(4) we plot the values of $\kappa$ vs. the corresponding
consistent value of input $|\mu|$.
We now find
the critical coupling, but with $\kappa_c= 6.82$, hence $g_0^2N_c/8\pi^2=1.082$,  
slightly larger
than
 the  NJL result, and very close to the exact Yukawa potential result $g_0^2N_c/8\pi^2 =1.06940$
 of eq.(\ref{exact}).  The tail-spline calculation has significantly improved the
 precision determination of the variational calculation of the critical behavior.
 
 Note that, 
 to an excellent approximation we now find a linear relation
of $\kappa$ and eigenvalue $\mu$ as 
\smallbbo
\label{linear}
\kappa = 6.8197+10.693|\mu|/M_0
\smallebo 
 Likewise,  $\phi(0)$ is then determined by $A$ in eq.(\ref{A}) to an excellent approximation
 in the limit eq.(\ref{tail}).
 
{
\begin{figure}
	\centering
	\hspace{0.1in}
	\includegraphics[width=0.58\textwidth]{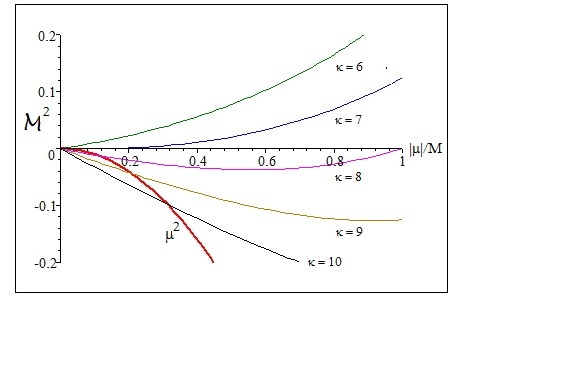}
	\vspace{-0.7in}
	\caption{ ${\cal{M}}^2$ of eq.(\ref{M}) with eq.(\ref{A}) is plotted vs. $\mu/M_0$ for $M_0=1$, 
	for values of $\kappa= (6, 7, 8, 9, 10)$. The thick (red) curve is the eigenvalue
	$\mu^2$. Consistent solutions occur where the $-|\mu|^2$ curve intersects the $\mu^2={\cal{M}}^2$ curves
	for given value of $\kappa$.
	The critical coupling is the smallest value of $\kappa$ for which these curves
	do not intersect, $\approx \kappa =6.8198$. For smaller values we have no solution
	with negative $\mu^2$.}
	\label{figvarfamily}
\end{figure}
}
{
\begin{figure}
	\centering
    \hspace{0.2in}
	\includegraphics[width=0.59\textwidth]{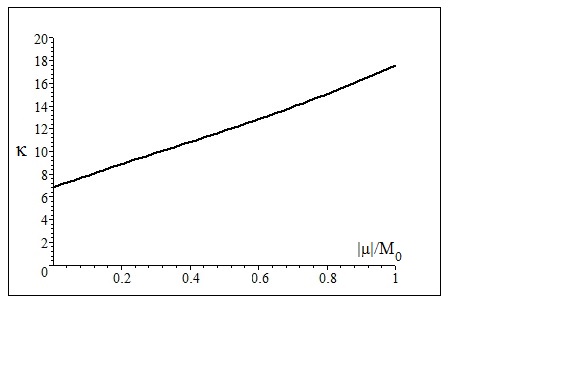}
  \vspace{-0.7in}
	\caption{ The value of the coupling $\kappa=g_0^2N_c/4\pi$ vs the bound state mass $|\mu|$ 
	which consistently matches the trial input value of $\mu/M_0$ to the eigenvalue $\mu^2= {\cal{M}}^2$
	(for negative $\mu^2$).  The result is fit by eq.(\ref{linear},
	$\kappa = 6.8197+10.693|\mu|/M_0$.
	This implies that the fine-tuning of 
	a hierarchy is significantly reduced as  $\delta \kappa/\kappa \sim |\mu|/M_0 $. }
	\label{figy}
\end{figure}
}

 If we again demand, $g_Y=1$, as in a  top quark condensation model, then from eqs.(\ref{gy},\ref{tail}),
 \smallbbo
\label{gylast}
g_Y= 1 = \hat{g}_Y\phi(0)
=\sqrt{\frac{N_c}{8}}\bl\frac{8\pi^2}{N_c}\br\frac{e}{3\sqrt{\pi} }\bl \frac{|\mu|}{M_0 }\br^{1/2}
\smallebo
hence
\smallbbo
\frac{ |\mu|}{M_0}=\bl\frac{N_c}{8\pi^2}\br^2\bl\frac{8}{N_c}\br\frac{9\pi}{e^2} =2.6\times 10^{-2} 
\smallebo\normalsize
with $N_c=3$ and the normalization at the origin.
In a top condensation model with $|\mu|=88$ GeV
this would imply a coloron mass scale of order $M_0\sim 6$ TeV. 
Due to the linear relationship between $\kappa$ and $|\mu|/M_0$
we see the degree of fine-tuning a hierarchy is of order
$\delta \kappa/\kappa \sim |\mu|/M_0 \sim 1.4\%$.
  This is  an astonishing improvement over the old NJL based 
   top condensate theory \cite{BHL}, an indicates a possible top quark
   coloron in the $\sim 5$--$10$ TeV range.

\section{Fermion Loops \label{loops}}

Presently we focus on the induced Yukawa interaction of eq.(\ref{extyuk}):
\smallbbo
\backo
S'
\rightarrow
-\sqrt{N_cJ/2} g^2_0M_0^2\times
\nonumbo\;\;
\!\int_{X\vvr}\!\![\bar{\psi}_L(X^+)\psi_{R}(X^-)]_{f}\chi(X)U(r) {+h.c.}
\smallebo\normalsize
where $X^{\mu\pm}=(X^0, \vvX \pm \vvr)$ and we define the combination,
\smallbbo
U(\vvr) = V(2\vvr)\phi(\vvr)
\smallebo\normalsize

This is the analogue of the fermion loop in the NJL model of figure(\ref{figvar})
as in the leading large-$N_c$ calculation. 
The Yukawa coupling involves $\phi(r)$ to 
the free fermion bilinear, $[\bar{\psi}_L(X^+)\psi_{R}(X^-)]$, 
via the potential $V(2r)$. 

The meaning of the ``free fermions''
now presents a subtlety: a free fermion bilinear
can itself be viewed as a bilocal field describing a scattering state,
$[\bar{\psi}_L(X^+)\psi_{R}(X^-)]\sim \Omega(X,r)$
and this would be a solution to the SKG equation with positive
eigenvalue, $q^2$, and must therefore be orthogonal to $\Phi(X,r)$: 
\smbbo
\!\int_{X\vvr}\!\!\Phi(X,r)^\dagger \Omega(X,r) =0
\smebo
This in turn implies a nontrivial constraint on the basis functions that comprise 
the free fermion states, and thus the Feynman propagator.
Essentially, we have extracted the particular combination of free fermion field basis functions
that comprise $\Phi$ from the complete set of basis functions that would
normally appear in an arbitrary $\Omega(X,r)$.

For the present calculation, however, we can consider $\Phi$ to be
a near critical bound state with an extended $\phi(r)$.  This serves
to dilute
the loop by a factor of $\phi(0)^2$. We would expect
that the orthogonality projection of $\Phi$ out of the set of free field 
basis functions would lead to a further correction of order $\phi(0)^4$.  
We will therefore approximate
the free fermions by plane wave Dirac fields and do the loop as
we would with the usual free Feynman propagator.

The free field Feynman fermion propagator then takes the usual form,
\smallbbo
\backo
S_F(x)=
i\int \frac{d^4\ell}{(2\pi)^4}\frac{\slash{\ell}}{{\ell^2}}e^{i\ell_\mu x^\mu}
\smallebo\normalsize
and the left handed projection operator is ${\cal{P}}_5=(1-\gamma^5)/2$
(we follow conventions of \cite{BjDrell}).

The  fermion loop
integral becomes, 
\smallbbo
\label{a122}
\backo
I= 
\nonumbo
\backo\backo
FM_0^4\!\!\int\!\!\! \frac{d^4\ell}{(2\pi)^4} d^3rd^3r'
\frac{\ell\!\cdot\!(\ell+\!{P})}{\ell^2(\ell+\!P)^2}
U(\vec{r})U^\dagger(\vec{r}')e^{2i{\vec{\ell}}\cdot(\vec{r}-\vec{r}')}
\smallebo
where we have the combinatorial overall factor,
\smallbbo
F\!=\!\ -2JN_c g_0^4\!N^2_c
\int d^4X|\chi_0|^2\!
\smallebo
Note the factor of $g_0^4N^2_c$ from the internal wave-function
interaction, eq.(\ref{extyuk}),
the additional factor of $N_c$ from the loop.

\subsection{$P=0$ Result}

For simplicity first consider $P=0$,
and  do the $\ell_0$ integral by
residues:
\smallbbo
\backo
\int\frac{d^4\ell}{(2\pi)^4}\frac{1}{\ell^2-\mu^2+i\epsilon}
=
\frac{i}{2}\int \frac{d^3\vec{\ell}}{(2\pi)^3}\frac{1}{(\vec{\ell}^2+\mu^2)^{1/2}}
\smallebo\normalsize
to obtain,
\smallbbo
\label{ten}
\backo
I=\frac{i}{2} F M_0^4\!\int\!\!d^3r d^3r'\int_\mu^{M_0}\!\!\!\!\frac{d^3\vec{\ell}}{(2\pi)^3}   
\frac{1}{|\vec\ell|}
U(\vec{r})U^\dagger(\vec{r}')e^{2i\vec{\ell}\cdot(\vec{r}-\vec{r}')}
\nonumbo 
 =
\frac{i}{32\pi^2 } F M_0^4 \int {d^3r} {d^3r'}
U(\vec{r})U^\dagger(\vec{r}')\times 
\nonumbo
\qquad
\times\left[\frac{\sin^2(M_0 |\vec{r}-\vec{r}'|)-\sin^2(\mu|\vec{r}-\vec{r}'|)}{|\vec{r}-\vec{r}'|^2}\!\right].
\smallebo\normalsize
where we introduced 
the $\vec{\ell}$ integral with UV cut-off $M_0$ and
IR cut-off $\mu$.

The result of eq.(\ref{ten}) is general. 
We can now take the pointlike  limit for the potential as in eq.(\ref{Ypot3})
to compare to the NJL model:
\smallbbo
V_0(2r)\rightarrow -\frac{1}{M^2_0}\delta^3(2\vvr)=-\frac{2}{JM^2_0}\delta^3(\vvr)
\smallebo\normalsize
to obtain:
\smallbbo
I=i\frac{N_c}{8\pi^2} |\phi(0)|^2(M_0^2-\mu^2) ( 2g_0^4N_c/J) 
\int d^4X|\chi_0|^2
\nonumbo
=i\frac{ \hat{g}^2_Y|\phi(0)|^2 N_c}{8\pi^2}(M_0^2-\mu^2) 
\int d^4X\!|\chi_0|^2
\smallebo\normalsize
Recall the induced Yukawa coupling of eq.(\ref{gy}),
\smallbbo
g_Y=   g_0^2\sqrt{2N_c/J}\phi(0) =\hat{g}_Y\phi(0)
 \smallebo\normalsize
to obtain,
\smallbbo
\backo I=
i{g}_Y^2N_c\int \! \! d^4X|\chi_0|^2 
\frac{M_0^2 -\mu^2}{8\pi^2}
\smallebo\normalsize
We have recovered
the usual momentum space results of eq.(\ref{NJL4}) as in \cite{BHL}.

Note the key difference with the NJL model is that, in NJL we would have ${g}_Y=g_0$, which is very large near criticality. 
However, presently we have 
$g_Y=\hat{g}_Y\phi(0)\equiv g_0^2\sqrt{2N_c/J}\phi(0)$ and
$g_Y$ is significantly
diluted near criticality by $\phi(0)$.

\subsection{$P$ Dependence}

Keeping the $P$ dependence to ${\cal{O}}(P^2)$ from eq.(\ref{a122}) yields,
$I=I_0+I_P$ and 
we use 
\smallbbo
\int\frac{d^4\ell}{(2\pi)^4}   
\frac{\ell\cdot(\ell+{P})}{\ell^2(\ell+P)^2}
=\frac{i}{2}\int \frac{d^3\vec{\ell}}{(2\pi)^3}
\bl
\frac{1}{|\vec{\ell}|}+
\frac{{P}^2}{2\hat\ell^3}\br
\smallebo
hence
\smallbbo 
I_P=
\nonumbo
\frac{i}{8} F M_0^4\!\!\int\!\!d^3r d^3r' \int_\mu^{M_0}\frac{ d|\vec{\ell}|}{4\pi^2}   
\frac{U(\vec{r})U^\dagger(\vec{r}')\sin(2|\vec{\ell}||\vec{r}-\vec{r}'|)}{|\vec\ell|^2|\vec{r}-\vec{r}'|}
\frac{P^2}{2}
\nonumbo
\smallebo\normalsize
We use an approximate result for the log-divergent integral,
\smallbbo
\backo
\int_\mu^{M_0} \frac{\sin(2xR)}{x^2R}dx 
\nonumbo
\backo\backo
\approx 
\frac{\sin(2\mu R)}{\mu R}\!-\!2(\gamma +\ln(2\mu R)-R^2\mu^2)+\!O\left(\!\frac{1}{M_0},\mu^3\right)
\smallebo\normalsize
to obtain,
\smallbbo I_P= \frac{i}{32\pi^2} F M_0^4\!\int\!\!d^3r d^3r' 
U(\vec{r})\U^\dagger(\vec{r}') \frac{P^2}{2}
\nonumbo
\bl \frac{\sin(2\mu |\vec{r}-\vec{r}'|)}{\mu|\vec{r}-\vec{r}'|}-2\ln(2\mu e^{\gamma} |\vec{r}-\vec{r}'|)  
-2\mu^2|\vec{r}-\vec{r}'|^2\br 
\nonumbo -(\mu \rightarrow  M_0) +{\cal{O}}(1/M_0,\mu^3)
\smallebo\normalsize
Taking the pointlike limit gives,
\smallbbo
\backo
I_P=i\frac{ \hat{g}^2_Y|\phi(0)|^2 N_c}{8\pi^2}P^2
\int d^4X|\chi_0|^2\!
\ln\bl\frac{2M_0}{\mu} e^{\gamma}  \br 
\nonumbo
=i\frac{ {g}^2_Y N_c}{8\pi^2}P^2
\int d^4X|\chi_0|^2\!
\ln\bl\frac{2M_0}{\mu} e^{\gamma}  \br 
 \smallebo\normalsize
where the log structure matches the result of eq.(\ref{NJL4}).
Hence the wave-function renormalization constant
of the $\Phi$ field is
\smallbbo
Z = 1+\frac{g_Y^2N_c}{8\pi^2} 
\bl \ln(M_0/\mu)  + c\br
\smallebo\normalsize
in agreement with the NJL loop result,
eq.(\ref{NJL4}) (the difference is
the additive constant in the limit $g_Y\rightarrow0$).
Hence this loop calculation, which drives the entire kinetic term structure of the
NJL model, is now a perturbative correction in our semiclassical scheme.

We remark that here we see something noteworthy: In the NJL model the limit
$|\vvr -\vvr{}'|< M_0^{-1}$ is inconsistent with treating $M_0$ as a momentum
space cut-off. That is, if we  insist upon momentum scales above $M_0$ to be disallowed
then in configuration space we must require  $|\vvr -\vvr{}'|\gta M_0^{-1}$. 
This informs us that the usual NJL model assumption of a cut-off
theory at scale $M_0$ is potentially inconsistent.

 \subsection{Quartic Interaction}
 
 As in the NJL model, the fermion loops will induce a quartic interaction.
A full calculation in configuration space is tedious but leads to results similar to
the following in the pointlike potential limit.
 
 The loop for a quartic interaction has
 four bilocal vertices and takes the form:
\smallbbo
\lambda =-\half F_4\int \frac{d^4\hat\ell}{(2\pi)^4}\frac{1}{\hat\ell^4}\displaystyle \prod_{i=1}^{4} d^3r_i   \;
|U(\vec{r}_i)|^4 e^{2i\hat{\vec{\ell}}\cdot\vec{r}_i}
\smallebo\normalsize
where
\smallbbo
\backo\backo
\int\!\! d^4r\; e^{2i\vvp_{1}\cdot \vvr} D(2r)\phi(\vvr)
=-\half \int\!\! d^3r\; e^{2i\vvp_{1}\cdot \vvr} V_0(2\vvr)\phi(\vvr)
\nonumbo
F_4\!=\!\bl N_c(\sqrt{2N_cJ} g^2_0M_0^2)^4
\int d^4X|\chi_0|^4\!\br
\smallebo
We integrate $\hat{\ell}^2$ from
$\mu^2$ to $M_0^2$ as IR and UV cut-offs (we could equally well include
$\mu^2$ in the propagator denominator for the IR cu-off  with similar results).
The pointlike limit for the potential is as above,
\smallbbo
V_0(2r)\rightarrow -\frac{2}{JM^2_0}\delta^3(\vvr)
\smallebo
and obtain 
\smallbbo
\backo
\frac{\lambda}{2} =F_4\bl \frac{1}{JM^2_0}\br^4 |\phi(0)|^4 \int_{\mu^2}^{M_0^2} \frac{d^4\hat\ell}{(2\pi)^4}
\frac{1}{\hat\ell^4}\int d^4X|\chi_0|^4\!
\nonumbo \backo
= \frac{N_c}{8\pi^2 }\hat{g}_Y^4|\phi(0)|^4 
\bl\ln \left(\!\! \frac{M_0}{\mu }\!\!\right)+{\cal{O}}\left(\!\frac{\mu^2}{M_0^2}\!\right) \br\int d^4X|\chi_0|^4 
\nonumbo
\smallebo

The log evolution matches the result for the pointlike NJL case 
of eq.(\ref{NJL4}), with $g_Y=\hat{g}_Y\phi(0)$,
\smallbbo
\label{quart1}
\lambda \approx \frac{N_{c}g_Y^{4}}{4\pi ^{2}}%
 \ln \left(\!\! \frac{M_0}{\mu }\!\!\right) +\makebox{constant.}
\smallebo

\section{Spontaneous Symmetry Breaking}

With super-critical coupling, $g_0>g_{0c}$, the bilocal field $\Phi(X,r)$ has a negative  (mass)$^2$
eigenvalue (tachyonic), $\mu^2$ 
with a well-defined localized wave-function.  In the region external to the
potential (forbidden zone)
the field is exponentially damped. At exact criticality there is a $1/r$
tail that switches to exponential damping for $g_0> g_{0c}$.
The supercritical solutions are localized and normalizable over the
entire space $\vvr$, but with $\mu^2<0$ they would lead to exponential
runaway in time of the field $\chi(X^0)$, and must be stabilized, typically
with a $\sim \lambda |\Phi|^4\rightarrow \lambda |\chi|^4$ interaction.

We then treat the supercritical case as resulting in spontaneous symmetry breaking. 
The spontaneous symmetry broken phase is then 
a configuration where the field $\Phi(X,r)=\chi(X)\phi(r)$, and
where $\chi(X)$ develops a vacuum expectation value (VEV), while $\phi(r)$ remains a 
localized wave-function satisfying the SKG equation. .
We then obtain  the ``sombrero potential'', 
  \smallbbo
 V(\chi) = -|\mu|^2|\chi|^2+\frac{\lambda}{2}|\chi|^4
  \smallebo\normalsize
The field $\chi$  develops a VEV, $\langle \chi \rangle =v =|\mu|/\sqrt{\lambda}$.

The external scattering state ``free'' fermions, $\psi_f^a(X)$, will then acquire mass 
through the  
Yukawa interaction, the second term in eq.(\ref{5NJL}),
$\sim g_Yv [\bar{\psi}_L\psi_R]_f + h.c. $.
 In our treatment the internal fermion pair belongs to $\phi(r)$
 which interacts with itself
 through the SKG equation and loop induced quartic interaction. It then
 drives the {\em external free fermions} to acquire mass. 
 At leading order there is no ``back reaction'' on the internal fermions 
 that bind to comprise $\phi(r)$.
 We segregated the free external fermions from the bound state
wave-function, $\Phi$, in eq.(\ref{5NJL}), so while 
$\Phi$ forms a VEV  as described above, and the scattering state fermions
independently 
acquire mass as spectators, the internal dynamics are as described by $\Phi_0$
as in eq.(\ref{5NJL}).

In a more general large-distance potential we may consider
possible feedback on the $r$ dependence of $\phi(r)$,
rather than the pointlike limit of the potential where things depend only upon $\phi(0)$.
The simplest large-distance sombrero potential might be modeled
as,
 \smallbbo
\backo\;
S=\!\!\int_{X\vvr}' \!\bl \!|\phi|^2\!\left|\frac{\partial\chi}{\partial X}\right|^2\!
\!\!-\!|\chi|^2( |\nabla_r\phi|^2\!
+\!g^2N_cM V(r)|\phi(r)|^2) 
\nonumbo
\qquad\qquad
-|\chi|^4\frac{\hat\lambda}{2}|\phi(\vvr)|^4\br
 \smallebo\normalsize
In the simplest case of a perturbatively small $\hat\lambda$
we expect the eigensolution of $\phi$
to be essentially unaffected
\smallbbo
\int'_r \!\! \bl \!-|\partial_{\vvr}\phi|^2\!
+\!{g^2N_c\!M}V(r)|\phi(\vvr)|^2\!\br \approx- \mu^2
\smallebo\normalsize
The effective quartic coupling is then further renormalized by
\smallbbo
|\chi|^4\int'_r \frac{\hat \lambda}{2}|\phi(\vvr)|^4=|\chi|^4\frac{\widetilde\lambda}{2}
\smallebo\normalsize
Then  $\chi$ develops a VEV in the usual way:
 \smallbbo
\langle |\chi|^2 \rangle = |\mu^2|/\widetilde\lambda =v^2
 \smallebo\normalsize
This is consequence of $\phi(\vvr)$ remaining localized in its potential

 However, for general $\widetilde\lambda$, possibly large
 (as in a nonlinear sigma model),
 the situation is potentially more complicated.
 The VEV is determined by joint integro-differential equations for constant $\chi$ 
 \smallbbo
 0= -\!\!\int'_r \bl \!-|\partial_{\vvr}\phi|^2\!
+\!{g^2N_c\!M}V(r)|\phi(\vvr)|^2\!
+{\widetilde\lambda}|\chi|^2||\phi(\vvr)|^4 \br
\nonumbo
0=
-\nabla^2_r\phi\!-\!{g^2N_c\!M}V(r)\phi(r)
-{\widetilde\lambda}|\chi|^2|\phi(r)|^2\phi(r)
 \smallebo\normalsize
 If we can can solve the second local equation then the global one
 follows, but we see that
  $\phi(r)$ cannot
 become constant in a potential $V(r)$ which has $r$ dependence! 
 While perturbative 
 solutions maintain locality in $\phi(\vvr)$, it is unclear what solutions exist to
 the latter equation for non-perturbative $\lambda$.

\section{Summary and Conclusions}

Bilocal fields, introduced by Yukawa \cite{Yukawa},  provided
a starting point for
a theory of correlated pairs of fermions in a Lorentz invariant action.
Our formalism, inspired by the NJL model \cite{NJL},
is a semiclassical theory
of bound states and 
yields a sensible physical picture. 
The introduction of the bilocal field, $\Phi(x,y)= \bar{\psi}_L(x)\psi_R(y)$,
is a bosonization of the fermion pair, and simplifies many
aspects of the formalism. We differ from Yukawa in the treatment of
``relative time.'' The physical mass scale for  a bound state
is inherited from the interaction. 
Our present construction leads to a number of
novel results and the following is a synopsis.

To describe a scalar bound state we can write a factorized bilocal field ansatz, 
$\Phi = \chi(X)\phi(r)$, where
$\chi(X)$ is  normal pointlike field that describes the center of mass motion, and $\phi(r)$
is the internal wave-function that describes the structure of the bound state.
In the rest frame, the relative time, $r^0$, disappears  and can be integrated out.
The ``internal wave-function'' $\phi(\vec{r})$ 
is then a static function of the constituent's separation, $2\vec{r}$, and $\chi(X)\rightarrow \chi(X^0)$
has only time dependence.

We consider the coloron model \cite{Bijnens}\cite{Topcolor}\cite{NSD}, 
which is a single boson exchange of
a massive gluon, of mass $M_0$, in leading order of large $N_c$.
This leads in the rest frame to a Hamiltonian for the internal
wave function with a static Yukawa potential:
\smallbbo
\label{final51}
\backo\!\!\!
{\cal{M}}^2
= M_0^3\!\!\int\!\!d^3r \bl |\partial_{\vec{r}}\phi |^2-g_0^2N_c M_0\frac{ e^{-2M_0 |\vvr|}}{8\pi |\vvr|}|\phi|^2\br 
\smallebo 
Here $\phi(r)$ is normalized as,
\smallbbo
M_0^3\int\!\! d^3 r\; |\phi(\vec{r})|^2=1
\smallebo
The potential is semiclassically enhanced by a factor of $N_c$, the number of colors,
by the color singlet normalization of $\Phi(x,y)$. 

By variation of ${\cal{M}}^2$ we obtain the Schr\"odinger-Klein-Gordon equation
and it's eigenvalue, 
 $\mu^2$:
  \smallbbo
\label{finalSKG}
\backo\backo
-\bl\frac{\partial^2 }{\partial r^2}+\frac{2}{r}\frac{\partial }{\partial r}\br\phi(r)\!
-\!g_0^2N_c M_0\frac{ e^{-2M_0 |\vvr|}}{8\pi |\vvr|}\phi(r) \! = \!\mu^2\!\phi(r)
 \smallebo\normalsize
 The eigenvalue, $\mu^2$, is then the physical {\em squared mass} of the bound state, $\chi(X)$, field in any frame.
Here ``binding'' represents a negative $\mu^2$ and a chiral instability of the vacuum. 

We also obtain an induced Yukawa coupling  of the bound state to
external unbound fermions,  $g_Y=g_0^2\phi(0)\sqrt{N_c/8}$. 
The main feature is that $g_Y\propto \phi(0)$ with
interesting consequences.

 We use the Hamiltonian directly in variational calculations of $\phi(r)$.
 First, we establish that the critical value of our Yukawa potential, i.e. the value 
 $g_0^2=g^2_{0c}$ for which $\mu^2=0$, 
  is equivalent to that determined in the literature for
 screened Coulomb potentials, (which are of the Yukawa form).  To high
 precision, \cite{Edwards} we find, remarkably, that 
 this corresponds very closely to the critical value
 in the NJL model obtained at loop level:
 \smallbbo
\label{finalexact}
\backo\!\!\!
\left.\frac{g_{0c}^2N_c}{8\pi^2}\right|_{screened}\!\!\!\! =1.06940
\qquad 
\left.\frac{g_{0c}^2N_c}{8\pi^2}\right|_{NJL}\!\!\!\!=1.00 .
\smallebo
 Moreover, in a  variational calculation,
 using a  splined-wave-function,
 we obtain in the present formalism:
 \smallbbo
\label{finalexact2}
\left.\frac{g_0^2N_c}{8\pi^2}\right|_{present}\!\!\!\!=1.082
\smallebo
in good agreement with the screened Coulomb result.  The quantitative agreement with
the loop level NJL model is striking and we are unaware of it's being previously noted.

For subcritical coupling
there are unstable resonances decaying
into their constituents, which are non-normalizable solutions of the SKG
equation with incoming and outgoing radiative tails.
Supercritical coupling, $g_0^2>g_{0c}^2$, implies a bound state 
with a negative $\mu^2$ eigenvalue, 
and therefore spontaneous chiral symmetry breaking must occur.

We mainly study the large $M_0$ limit in which the Yukawa
potential approaches a $\delta^3(\vec{r})$ potential.  While this
becomes the NJL potential, nonetheless, 
in this limit $\phi(r) $ remains a spatially extended field 
and the theory remains non-pointlike. This is  a major
difference with the NJL model.  

Our most important results have to do with $\phi(r)$ near criticality.
As we approach critical coupling $g_0^2\rightarrow g_{0c}^2$
the $\phi(r)$ becomes scale invariant, and 
we have for $r>\!\!>M_0^{-1}$,
\smallbbo 
\phi(r)\sim  \frac{A e^{-|\mu| r}}{r}\rightarrow  \frac{A}{r} \qquad  |\mu| <\!\!< M_0 
\smallebo
where $A$ is the normalization constant.
The normalization of $\phi(r)$ is then dominated by
the long distance tail. 
As $r\rightarrow 0$, and $|\mu|<\!\! M_0$, the $\phi(r)$
solutions tend to a constant, $\phi(0)$.  This is suppressed
as $\phi(0) \propto (|\mu|/M_0)^{1/2}$.

This ``infrared dilution effect'' of $\phi(0)<\!\!< 1$ 
has significant implications on the results of the theory that
are quite different than those of the NJL model.
The induced Yukawa coupling is  $g_Y\propto \phi(0)$.
This means that, though $g_0^2 >\!\!>1$, the value of
$g_Y<\!\!< g_0$ dilutes quickly to small values.  
In contrast, in the NJL model the value of $g_Y$ is 
suppressed, but only logarithmically $g_Y\sim 1/\ln(M_0/\mu)$ at leading
$N_c$ fermion loop level,  and evolves relatively
slowly with scale into the IR, even if the full renormalization group
is applied \cite{BHL}\cite{PR}.
In the present semiclassical scheme, the  suppression
is fast and power-law, $\sim (|\mu|/M_0)^{1/2}$.  This decouples
the strong dynamics underlying the bound state at short distances, making it
effectively perturbative at low energies.  The
 ``custodial symmetry'' of this dilution is scale invariance, as
 we  approach the critical coupling.

For example, applying the NJL model in top condensation models \cite{BHL}\cite{Topcolor}\cite{NSD},
then $\mu^2=-(88)^2$ GeV$^2$, the Lagrangian BEH mass of the standard model, and
we would typically require $|\mu|/M_0\sim 10^{-15}-10^{-19}$  to get
$g_Y$ of order unity ($g_Y$ never reaches unity and tends to the  
IR fixed point value \cite{PR}).
In the present semiclassical scheme, owing to the $\phi(0)$ dilution,
one can achieve $g_Y=1$ with $M_0\sim 6$ TeV! 

Moreover, the critical behavior is significantly modified.
Typically, as in the NJL model, the critical behavior would go as
a second order phase transition where,
\smallbbo
\label{fine}
\mu^2 = M_0^2 - \frac{g_0^2}{g_{0c}^2}M_0^2
\smallebo
This implies significant fine-tuning to obtain a hierarchy, 
 at the level of  $(|\mu|/M_0)^2 \sim 10^{-28}$, or more, in the NJL 
 top condensation scheme.
However, in the present framework the {\em rhs} of eq.(\ref{fine}) is renormalized by
$\phi^2(0)$, and we obtain,
\smallbbo
\label{fine2}
\backo
\mu^2 = \phi^2(0)\bl M_0^2 - \frac{g_0^2}{g_{0c}^2}M_0^2\br
=  |\mu| {M_0}\bl 1 - \frac{g_0^2}{g_{0c}^2}\br
\smallebo
Thus we  have a linear relationship between $g_0^2$ and $|\mu|$,
\smallbbo
\label{fine2}
\frac{|\mu|}{M_0} 
= \bl \frac{g_0^2}{g_{0c}^2}-1\br, \qquad g_0^2>g_{0c}^2.
\smallebo
This implies that fine-tuning a hierarchy
is now significantly reduced to
$\delta g_0^2/g_0^2 \sim \mu/M_0 \sim 10^{-2}$... a few \%
with $M_0\sim 6$ TeV!

Also, in the NJL model of top condensation  the value of the quartic coupling
is determined by the renormalization group with ``compositeness boundary conditions''
and is generally too large (this is problematic for many 
theories of a composite BEH boson).
However, owing to the suppression of $g_Y$ the quartic coupling
is now generated in loops as in eq.(\ref{quart1}) and found to be close to the standard model result.

We think this bodes well for a renaissance of the top condensation/topcolor 
scheme, or perhaps  other constituent  models of a composite BEH boson.
We will revisit this  elsewhere \cite{CTHprep}.
We believe the most important challenge to the LHC is to ascertain whether
or not the BEH boson is pointlike or an extended object (e.g. showing deviations
from the standard model, particularly in 3rd generation processes, or perhaps
by way of tools such as
\cite{Burdman} and others).

In summary, our key result is that, near criticality, the NJL model fails, while a semiclassical theory contains additional degrees of freedom, i.e., an internal wave-function
$\phi(r)$, and the major low energy results are significantly modified. Near criticality
the low energy $\phi(r)$ is approximately dynamically scale invariant,
and scale symmetry is acting as the custodial symmetry of
the physics of a low mass bound state
of chiral fermions.
Our results may be of some general interest to practitioners of the NJL
model in QCD, e.g., \cite{Bijnens},\cite{NJLReview}, etc.
For example we think it would be interesting to apply this formalism to
heavy-light systems as in \cite{bardeenhill}.

We remark that
the transition from unbound to bound state is associated
with the internal wave-function becoming a ``compact''  solution
to the SKG equation. Just below
critical coupling, $g_0<g_{0c}$, the
eigenfunctions at large $r$ are two body scattering states, such as $\sim e^{i q r}/r$,
requiring space-volume, $V_3$, normalization, $\sim 1/\sqrt{V_3}$. 
 For $g_0>g_{0c}$ the internal bound state eigenfunction discontinuously
becomes compact and normalizable $\sim e^{-\mu r}/r$.
This transition is non-analytic in momentum space, but intuitive,
in configuration space as reflected in the eigenvalue flow
from subcritical to critical coupling.

Weinberg has emphasized the non-analyticity of the transition from unbound
to bound state in momentum space as an outstanding challenge to bound state formalism \cite{Weinberg}. Tree level
scattering amplitudes for unbound pairs
are perturbative, and can be approximated to any desired order by a {\em finite number}
of Feynman
diagrams, but generate no bound state.  A bound state requires summing an {\em infinite
number} of loop diagrams to generate the bound state pole. In practice this involves a choice of particular subset
of diagrams to sum, e.g., in a Bethe-Salpeter equation or a fermion loop ``bubble sum'' approximation to
the NJL model.  However,
at this stage the procedure can 
become non-systematic (except in a well defined subset, e.g., large-$N_c$ fermion loop approximation in NJL).
Therefore, the usual diagrammatic expansion in the coupling
has a discontinuity, a non-analytic behavior, as binding commences.

This non-analyticity in momentum space traces in configuration space
to the transition from non-normalizable
scattering state wave-functions to the normalizable bound state wave-functions.
In the present formalism the transition is
a first order phase transition in 
$\epsilon=0$ to $\epsilon=1$  as $g^2_0$ crosses from subcritical to critical.
Our present configuration space framework therefore offers, perhaps, a more intuitive view
of the bound state in field theory.
We can envision more applications of this approach. 

\appendix

\section{Review of the Point-like Nambu--Jona-Lasinio Model \label{NJLreview} }

The Nambu--Jona-Lasinio model (NJL) \cite{NJL} is the simplest field
theory of a composite scalar boson,
consisting of chiral fermions. 
An effective {\em pointlike} bound state emerges 
from an assumed {\em pointlike} 4-fermion interaction.
We begin with a lightning review of the modern solution
of the NJL model using concepts of the renormalization group.

We assume chiral fermions, each with $N_c$ ``colors.''
A non-confining,  chiral invariant  $U(1)_{L}\times U(1)_{R}$
action, then takes the form:
\bea &&
\label{NJL1}
S_{NJL} 
=\!\int\! d^4x \;\bl i[\bar{\psi}_L(x)\slash{\partial}\psi_{L}(x)]
+ i[\bar{\psi}_R(x)\slash{\partial}\psi_{R}(x)]
\nonumber \\ &&
\qquad \qquad
+\;
\frac{g_0^2}{M_0^2}
[\bar{\psi}_L(x)\psi_{R}(x)]\;[\bar{\psi}_R(x)\psi_{L}(x)]
\br.
\eea
Here $\psi_L = (1-\gamma_5)\psi/2$,
$\psi_R = (1+\gamma_5)\psi/2$, and we write color singlet 
combinations in brackets $[...]$ 
as $\bar{\psi}^i_L(x)\psi_{iR}(x)
\equiv [ \bar{\psi}_L(x)\psi_{R}(x)]$.
This can be readily generalized to a $G_L\times G_R$ chiral symmetry
which we do explicitly in the section on currents below.

We can rewrite eq.(\ref{NJL1}) in an equivalent form
by introducing the local color singlet auxiliary
field $\Phi(x)$:
\smallbbo
\label{NJL2}
\backo
S_{NJL}
=\!\int\! d^4x \;\bl i[\bar{\psi}_L(x)\slash{\partial}\psi_{L}(x)]+ i[\bar{\psi}_R(x)\slash{\partial}\psi_{R}(x)]
\nonumbo
- M_0^2\Phi^\dagger(x) \Phi(x) + g_0[\bar{\psi}_L(x)\psi_{R}(x)]\Phi(x)+h.c.  \br.
\smallebo
\normalsize  
The resulting ``equation of motion'' for $\Phi$ is:
\smallbbo
\label{NJL222}
 M_0^2\Phi(x) = g_0[\bar{\psi}_R(x)\psi_{L}(x)]
\smallebo\normalsize 
Using the $\Phi$ equation of motion in eq.(\ref{NJL2}) 
reproduces  the 4-fermion interaction of eq.(\ref{NJL1}).

Note that the induced (unrenormalized) 
Yukawa coupling $g_0$ in eq.(\ref{NJL2}) is the same coupling as
appears in the interaction of eq.(\ref{NJL1}).  In the semiclassical treatment
of this paper this relationship is significantly modified as in eq.(\ref{gy}).

{
\begin{figure}
	\centering
	\includegraphics[width=0.4\textwidth]{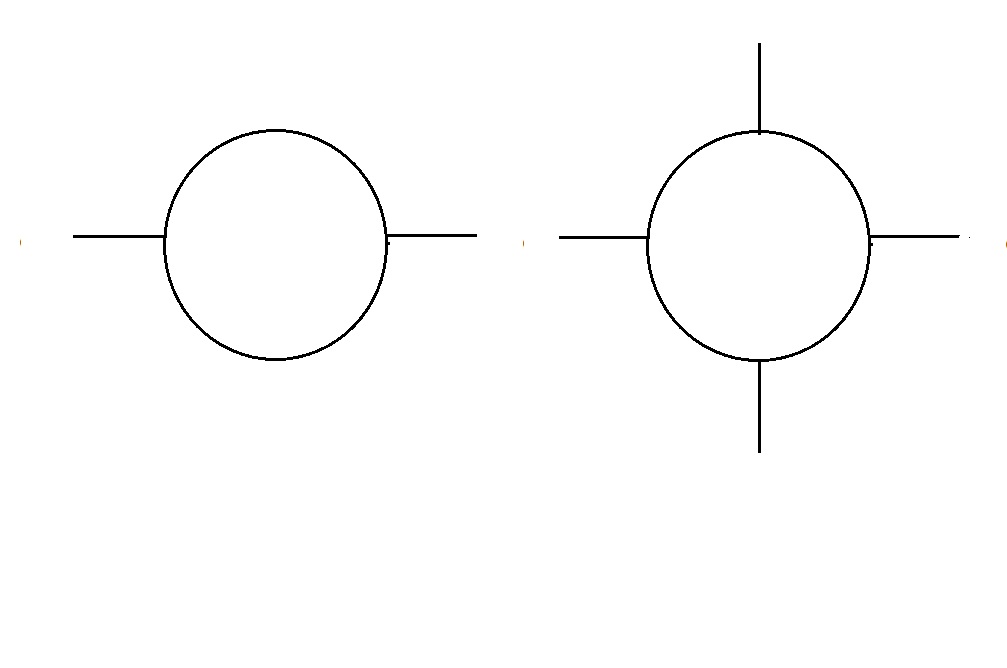}
	\vspace{-0.6in}
	\caption{ Diagrams contributing to the pointlike NJL model effective
	Lagrangian, eqs.(\ref{NJL5}). External lines are $\Phi$ and internal
	lines are fermions $\psi$.}
	\label{figvar}
\end{figure}
}

Following Wilson, \cite{Wilson},
we view eq.(\ref{NJL2}) as the action defined at the high energy (short-distance) scale $m \sim M$.
We then integrate out the fermions to obtain the effective action for the composite 
field $\Phi$ at a lower scale $m<\!\!<M$. 
The calculation in the large-$N_c$ limit
is discussed in detail in \cite{BHL,CTH}.  
The leading $N_c$ fermion loop yields the result
for the $\Phi$ terms in the action at a new scale $\mu$:
\smallbbo
\label{NJL3}
\backo
S_{\mu}
=\!\int\! d^4x \;\bl i[\bar{\psi}_L\slash{\partial}\psi_{L}]+ i[\bar{\psi}_R\slash{\partial}\psi_{R}]
+
Z\partial_\mu \Phi^\dagger\partial^\mu \Phi
\nonumbo
\backo
-\mu^2\Phi^\dagger \Phi - \frac{\lambda }{2}(\Phi^\dagger \Phi)^2 + g_0[\bar{\psi}_L\psi_{R}]\Phi(x)+h.c. )
\br.
\smallebo\normalsize 
where the diagrams of Fig.(1) yield,
\smallbbo
\label{NJL4}
\backo
\mu^2 = M_0^{2}\!-\!\frac{N_{c}g_0^{2}}{8\pi ^{2}} M_0^{2},
\nonumbo
\backo
Z=\frac{N_{c}g_0^{2}}{8\pi ^{2}}\ln(M_0/\mu), \;\;\;
\lambda=\frac{N_{c}g_0^{4}}{4\pi ^{2}}\ln( M_0/\mu).
 \smallebo\normalsize
Here $M_0^2$ is the UV loop momentum 
cut-off, and we include the induced kinetic and quartic interaction terms.
The one-loop result can be improved by using the full renormalization group (RG) \cite{BHL,CTH}.
Hence the NJL model is driven by fermion loops, which are $\propto \hbar$ intrinsically
quantum effects.

Note the behavior of the composite scalar boson mass, 
$\mu^2$, of eq.(\ref{NJL4}) in the UV.
The $ -{N_{c}g_0^{2}M_0^{2}}/{4\pi ^{2}}$  term arises from the negative
quadratic divergence in the loop involving the
pair $\left( {\psi }_{R},\psi _{L}\right) $ of Fig.(1), with
 UV cut-off  $M_0^{2}$. 
Therefore, the NJL model has a critical
value of its coupling defined by the cancellation of the large $M_0^2$ terms,
\smallbbo
g_{0c}^{2}=\frac{8\pi ^{2}}{N_{c}}
\smallebo\normalsize
Here we have  $m$ as the running RG mass, and is the lower limit
of the loop integrals. This can in principle be small in logs and neglected in
the quadratically divergent loops. 
At the critical coupling the mass of the bound state is then $\mu^2= 0$.

We can renormalize, $\Phi\rightarrow \sqrt{Z}^{-1}\Phi$, hence:
\smallbbo
\label{NJL5}
\backo
S_{\mu}
=\!\int\! d^4x \;\bl i\bar{\psi}^a_L\slash{\partial}\psi_{aL}+ i\bar{\psi}^a_R\slash{\partial}\psi_{aR}
+
\partial_\mu \Phi^\dagger\partial^\mu \Phi
\nonumbo
\backo
-\mu_r^2\Phi^\dagger \Phi - \frac{\lambda_r }{2}(\Phi^\dagger \Phi)^2 + g_Y\bar{\psi}^a_L\psi_{aR}\Phi(x)+h.c. 
\br.
\smallebo\normalsize 
where,
\smallbbo
\label{NJL30}
\backo
\mu_r^2 = \frac{1}{Z}\bl M_0^{2}\!-\!\frac{N_{c}g^{2}}{8\pi ^{2}} ( M_0^{2})\br
\nonumbo
\backo
g^2_Y \!=\! \frac{g^2}{Z}\!=\!\frac{8\pi^2}{N_c\ln(M_0/m)},\;\;
\lambda_r\!=\!\frac{\lambda}{Z^2}\! =\!\frac{16\pi ^{2}}{N_c\ln( M_0/m)}.
 \smallebo\normalsize

For  super-critical coupling, $g_0^2>g^2_{0c}$,
we see that $\mu_r^2<0$ and there will be a chiral vacuum instability.
The effective action, with the induced quartic  $\sim \lambda_r(\Phi^\dagger\Phi)^2$ term, 
is then the usual sombrero potential.
The chiral symmetry is spontaneously broken, the field $\Phi$ acquires a VEV,
\smallbbo
\langle\Phi\rangle =v = \frac{\mu_r}{\sqrt{\lambda_r}},
\smallebo\normalsize
and the
chiral fermions acquire mass, 
\smallbbo
m_f=g_Y v
\smallebo\normalsize
The physical radial mode (``Higgs'' boson), defined as $ \sqrt{2} Re(\Phi)$, has a mass $m_h$
given by,
\smallbbo 
m_h^2=2\lambda v^2.
\smallebo\normalsize 
The Nambu-Goldstone mode, $Im(\Phi)$, is massless.
Hence we see that the NJL model yields a prediction
for the radial mode
\smallbbo
m_h = \frac{\sqrt{2\lambda_r}}{g_Y}m_f=2m_f .
\smallebo\normalsize
The effective action also generates 
Nambu-Goldstone bosons that also emerge
as pointlike bound states.

Fine-tuning of $g_0^2 \rightarrow g_{0c}^2$ can be done 
to attempt to create  a hierarchy, $|\mu^2|<\!\!<M_0^2$.
In that case we appeal to the behavior of the renormalized 
couplings as $\mu\rightarrow M_0$. We see that both $g_Y$
and $\lambda_r$ diverge in the ratio $\lambda_r/g^2_Y\rightarrow 2$.
This can be used as a boundary condition on the full
RG evolution of $g_Y$ and $\lambda_r$
including gauge fields and scalar interactions themselves.

The NJL model is useful in QCD applications and 
supercritical coupling.
However, in the present semiclassical treatment of binding we claim
that the results of the critical coupling limit of the NJL model 
are incorrect. The model does not 
include the internal wave-function, $\phi(r)$, which behaves
as $\phi(r)\propto e^{-\mu r}/r\sim 1/r$ in the limit, and 
dilutes $g_Y\propto \phi(0)\sim \sqrt{\mu M_0}$ by a power-law.

 \section{Currents \label{currentsapp}}
 
  Under a $G_L\times G_R$ chiral symmetry transformation we have for free fermions,
 \smallbbo
\psi_L(y) \rightarrow G_{L} \psi_{L}(y)
\nonumbo
\psi_{R}(y)\rightarrow G_{R}{\psi}_{R}(y)
\smallebo
\normalsize
where dotted indices refer to $G_L$ and undotted to $G_R$.
  Therefore the free fermion field theory has chiral currents,
 \smallbbo
 \backo
 j^A_{L\mu}=[\bar{\psi}_L(x)\gamma_\mu T^{A}_L\psi_L(x)]
 \nonumbo
 \backo
 j^A_{R\mu}= [\bar{\psi}^a_R(x)\gamma_\mu T^{A}_R{}\psi_R(x)]
 \smallebo
 \normalsize
 where $[...]$ means we have contracted color indices and $T_L$ ($T_R$)
 is the generator of $G_L$ ($G_R$).
 These groups have corresponding currents in the composite two body field $\Phi$.
 The defining equation for $\Phi(x,y)$ implies its chiral symmetry properties
 under $G_L\times G_R$,
  \smallbbo
 \label{Phi100}
M_0^2\Phi(x,y)\rightarrow 
G_{L}\Phi(x,y)G^\dagger_{R}
 \smallebo

 Conserved currents provide a connection between the normalization
 of the fundamental constituent fields and the composite fields.
 The current normalization of $\Phi$ is equivalent to it's kinetic term
 normalization.
  The values of the associated charges lock the fundamental fermion
 fields, $ \psi_R(x) $  and $ \psi_L(y)$
  to the composite field $\Phi(x,y)$. 
 The matching of the composite to the constituent
 currents can be made exact for scalar constituents, which we do in \cite{bilocal}.
 The matching will also hold for fermions as operator constraints in the two body sector
 of the Fock space of states, as is the case for chiral Lagrangians in general.
 
 It is useful to focus on the {\em global} $U(1)_L\times U(1)_R$ subgroup which
 s present for any $G_L\times G_R$ generalized chiral symmetry group.
 With the two currents,
  \smallbbo
 \backo\!\!\!
 j_{L\mu}=[\bar{\psi}_L(x)\gamma_\mu \psi_{L}(x)]
 \qquad 
 j_{R\mu}= [\bar{\psi}_R(x)\gamma_\mu \psi_{R}(x)]
 \smallebo\normalsize
We have the vector current $j=j_L+j_R$ and an axial vector
 $j^5= j_L-j_R$, corresponding to $U_V(1)\times U_A(1)$

The symmetries act upon the  $\Phi(x,y)$  as a local gauge transformation
 of the form, 
 \smallbbo
 \Phi(x,y)\rightarrow U_R^\dagger(x)\Phi(x,y)U_L(y)
 \smallebo\normalsize
 where in principle the gauge rotations
 are different at $x$ and $y\neq x$.  In this way we can
 introduce chiral gauge fields.
 If, however, the gauge transformation is global,
 then the $U$'s are independent of $x,y$. 
 In barycentric coordinates we have  
  $U(X+r)=U(X-r)=U(X)$
 hence in the factorization form $\Phi'(X,r)\sim \chi(X)\phi(r)$
 we can assign the global symmetry representation to
 $\chi$
 and treat $\phi(r)$ as a scalar.   However, we require
 local transformations to generate Noether currents.
 Hence,
\smallbbo
 \Phi(x,y)\rightarrow e^{-i\theta_L(x}\Phi(x,y),
 \nonumbo
 \Phi'(X,r)\rightarrow e^{i\theta_R(y)}\Phi(x,y)
 \smallebo\normalsize
 generates the bilocal currents from eq.(\ref{xxNJL0}),
  \smallbbo
 J_{L\mu}(x,y)
 =iZM^4\Phi^\dagger(x,y)) 
 \frac{ \stackrel{\leftrightarrow}{\partial} }{ \partial x^\mu }\Phi(x,y)
 \nonumbo
 J_{R\mu}(X,r)
 =iZM^4\Phi^\dagger(x,y)) 
 \frac{ \stackrel{\leftrightarrow}{\partial} }{ \partial y^\mu }\Phi(x,y)
 \smallebo\normalsize
 
  Likewise, jumping to the $\Phi'(X,r)$ representation and
 the bilocal action eq.(\ref{af2}), we generate two Noether currents corresponding
 to $U^+(1)\times U^-(1)$, via the local 
 transformations,\footnote{Note that the constraint takes the form
  \smbbo
  \backo
 |\Omega|^2=\bl\frac{\partial \chi^\dagger(X)}{\partial X^\mu} \frac{\partial \phi(r)}{\partial r^\mu}
 \phi^\dagger(r) \chi(X)+h.c.\br^2=0
 \smebo
 The constraint is invariant under the global $U(1)_L\times U(1)_R$ transformations.
 Under the local transformation it generates 
 \smbbo
 \backo\!\!
 \bl i\frac{\partial\theta^+(X)}{\partial X^\mu} \frac{\partial \phi(r)}{\partial r_\mu}
 \phi^\dagger(r) |\chi(X)|^2+h.c.\br\Omega+h.c.=0
 \nonumbo
 \backo\!\!
 \bl -i\frac{\partial\chi^\dagger(X)}{\partial X^\mu} \frac{\partial\theta^-(r)}{\partial r_\mu}
 |\phi(r)|^2 \chi(X)+h.c.\br\Omega+h.c.=0
 \smebo
}
 
\smallbbo
 \Phi'(X,r)\rightarrow e^{i\theta^+(X)}\Phi'(X,r),
 \nonumbo
 \Phi'(X,r)\rightarrow e^{i\theta^-(r)}\Phi'(X,r)
 \smallebo\normalsize
 The bilocal currents are,
  \smallbbo
 J^+_{\mu}(X,r)
 =iZ'M^4[\chi^\dagger(X) 
 \frac{ \stackrel{\leftrightarrow}{\partial} }{ \partial X^\mu }\chi(X)]\phi^\dagger(r)\phi(r)
 \nonumbo
 J^-_{\mu}(X,r)
 =iZ'M^4[\phi^\dagger(r) 
 \frac{\stackrel{\leftrightarrow}{\partial}}{\partial {r^\mu}}\phi(r)]\chi^\dagger(X)\chi(X)
 \smallebo\normalsize
 These can be integrated to form, 
 \smallbbo
 J^+_{\mu}(X)
 =iZM^4[\chi^\dagger(X) 
 \frac{ \stackrel{\leftrightarrow}{\partial} }{ \partial X^\mu }\chi(X)]\int_r \phi^\dagger(r)\phi(r)
 \nonumbo
 J^-_{\mu}(r)
 =iZM^4[\phi^\dagger(r) 
 \frac{\stackrel{\leftrightarrow}{\partial}}{\partial {r^\mu}}\phi(r)]\!\int_X \!\!\chi^\dagger(X)\chi(X)
 \smallebo\normalsize
 The charge corresponding to $J^+$ counts the number of $\psi_L$ plus $\psi_R$
 particles and we can therefore match $J^+_{\mu}$ 
 to the underlying $j_V=j_1+j_2$ vector current.
 With eq.(\ref{Znorm}) and eq.(\ref{phinorm}) we have 
 \smallbbo
 1=ZM^4\int\!\! d^4r|\phi(r)|^2=M^3\int\!\! d^3r|\phi(r)|^2
 \smallebo\normalsize
 and the $U^+(1)$ current becomes,
 \smallbbo
 J_{\mu}(X)
 ={i}\chi^\dagger(X) 
 \frac{ \stackrel{\leftrightarrow}{\partial} }{ \partial X^\mu }\chi(X)
 \smallebo\normalsize

 If we consider a pointlike field $\Phi(X)$ then
 we lose the independent $U_A(1)=j_R-j_L$ transformation,
 i.e. $e^{i\theta(r)}$ is meaningless for the local field, and $\Phi(X)$
 can only represent a single $U_V(1)$ transformation.
 We are using the term
 ``(+)'' current because $\Phi\rightarrow e^{i\theta(X)}\Phi(X)=
 e^{i\theta(X)}\Phi(X)$ can correspond to either a vectorlike (electromagnetic) gauge transformation
 or a $U^5(1)$ axial transformation. The concept of parity arises 
 in the bosonic chiral Lagrangian only when we include
 couplings to chiral fermions.
 E.g., with a coupling to fermions,  $\sim\bar{\psi}_L\Phi\psi_R$
 the $U^+(1)$ becomes the axial transformation $\psi_L\rightarrow e^{-i\theta}\psi_L$,
 $\psi_R\rightarrow e^{i\theta}\psi_R$.
 In the pointlike limit where  $\Phi$ is invariant under
 $\Phi\rightarrow e^{i\theta(X)}\Phi(X)e^{-i\theta(X)}$ the corresponding
 $U^-(1)$ current is zero.
 However, the bilocal case has a nonzero $U^-(1)$ current.

 However, if we have a static $\phi(\vvr)$
 then the axial charge is zero,
 \smallbbo
 \backo\!\!
 J^-_{0}(X,r)
 =iZ[\phi^\dagger(\vvr) 
 \frac{\stackrel{\leftrightarrow}{\partial}}{\partial {r^0}}\phi(\vvr)]\int_X \chi^\dagger(X)\chi(X)
 =0
 \smallebo\normalsize
 
 Note that we can define a phase in analogy to a would be $U_5(1)$ Nambu-Goldstone boson,
 such as the $\eta'$,
 \smallbbo
 \phi(r)= e^{i\eta'(r^0)/f}\phi(\vvr)
 \smallebo\normalsize
 and the static condition implies in the rest frame,
 \smallbbo
 \partial_r^0\eta'(r^0)=0 \qquad \eta'=\theta=\makebox{constant}.
 \smallebo\normalsize
 This is suggestive of a mechanism to elevate the mass of the $\eta'$
 such as instantons and the $\eta'$ is therefore a constant at the minimum of a
 deep potential. While any other Nambu-Goldstone boson,
 such as the $\pi$, can have zero momentum but nonzero time dependence, 
 the $\eta'$ cannot be a Nambu-Goldstone boson if the constraint
 is applied.

\section{Colorons and Fierz Rearrangement$\label{spinor}$ \label{Fierz}}

\normalsize

The coloron model is a massive, perturbative gluon field, $B_A^{\mu }$, in
an $SU(N_c)$ gauge theory broken to the diagonal global  $SU(N_c)$. We assume fermions 
couple through vector color currents and a coloron mass $M_0$, with relevant
Lagrangian terms:
\smallbbo 
-g_0 \bar{\psi}\!(x)\!\gamma_\mu T^A \psi  B_A^{\mu } +M_0^2 B_A^{\mu}B_{A\mu}
\smallebo\normalsize
Integrating out the coloron yields the interaction,
\smallbbo
\label{TC0app}
\backo
S'\!=\!
\frac{1}{2}\!\!\int_{xy}\!\!\! 
 [-ig_0 \bar{\psi}\!\gamma_\mu\! T^A \psi]_x\langle T B_A^{\mu}(x) B_B^{\nu}(y)\rangle
[-ig_0\bar{\psi}{}\! \gamma_\nu\! T^B \psi]_y
\nonumbo
\backo
=-\frac{g_0^2}{2}\!\!\int_{xy}\!\!\! 
 [\bar{\psi}\!(x)\!\gamma_\mu T^A \psi\!(x)]D^{\mu\nu}(x-y)
[\bar{\psi}{}(y)\! \gamma_\nu T^A \psi(y)]
\smallebo
where we note that $iD^{\mu\nu}(x-y)\delta^{AB} =\langle T\; B^A_\mu(x) B^B_\nu(y)  \rangle $)
and we strip off a factor of $+i$
for the action. In Feynman gauge \cite{BjDrell}, 
\smallbbo
D_{\mu\nu}=g_{\mu\nu}D_F(x-y)
\nonumbo
D_F(x-y)=-\int \frac{d^4q}{(2\pi)^4}\frac{e^{-iq(x-y)}}{q^2-M_0^2+i\epsilon}
\smallebo

We are interested in chiral fermions and write $\psi=\psi_L +\psi_R$
with chiral projections
\smallbbo
\psi_L=\frac{1-\gamma^5}{2}\psi \qquad \psi_R=\frac{1+\gamma^5}{2}\psi 
\smallebo
and the interaction of interest becomes the cross-term of $L$ and $R$
currents in eq.(\ref{TC0}):
\smallbbo
\label{TC0app}
\backo
\;S'\!=\!
{-g_0^2}\!\!\int_{xy}\!\!\! 
 [\bar{\psi}_{L}\!(x)\!\gamma_\mu T^A \psi_{L}\!(x)]D^{\mu\nu}(x-y)
[\bar{\psi}{}_{R}(y)\! \gamma_\nu T^A \psi_{R}(y)]
\nonumbo
\smallebo
where $[...]$ denotes color summed indices, e.g.
\smallbbo
\backo\backo
[\bar{\psi}_{L}\!\gamma_\mu\psi_{L}] \equiv \bar{\psi}^i_{L}\!\gamma_\mu \psi_{iL},
\;\;\;
[\bar{\psi}_{L}\!\gamma_\mu T^A \psi_{L}]\equiv \bar{\psi}^i_{L}\!\gamma_\mu T^{A}{}^j_i \psi_{jL}
.
\smallebo
The interaction can be Fierz transposed. We define the operators
for generic fermions, $\psi_i$, sequentially ordered as $(1234)$:
\smallbbo
\backo\backo
{\cal{O}}_1= \bar{\psi}_1\psi_2\bar{\psi}_3\psi_4 ,  \;\;\;\;\;
{\cal{O}}_2= \bar{\psi}_1\gamma_\mu \psi_2\bar{\psi}_3\gamma^\mu\psi_4 
\nonumbo \backo\backo
{\cal{O}}_3= \bar{\psi}_1\sigma_{\mu\nu} \psi_2\bar{\psi}_3\sigma^{\mu\nu}\psi_4 
\nonumbo  \backo\backo
{\cal{O}}_4= \bar{\psi}_1\gamma^5\gamma_\mu \psi_2\bar{\psi}_3\gamma^5\gamma^\mu\psi_4, \;\;\;\;\;
{\cal{O}}_5= \bar{\psi}_1\gamma^5\psi_2\bar{\psi}_3\gamma^5\psi_4   
\smallebo
and define operators $\bar{\cal{O}}_i$ identical to the above
but  reordered as $(1432)$.  Note that we can allow any given fermion
to have chiral projection, e.g. $\psi_1 \rightarrow \psi_{1L}$, etc.
We then have the Fierz identity:
\smallbbo
{\cal{O}}_i= \displaystyle\sum_{j} M_{ij}\bar{\cal{O}}_j
\smallebo
where the matrix is,
\smallbbo
\label{fmatrix}
M_{ij}=-\frac{1}{8}\begin{pmatrix}
   2 & 2 & 1 & -2 & 2 \\
   8 & -4 & 0 & -4 & -8 \\
   24 & 0 & -4 & 0 & 24 \\
   -8 & -4 & 0 & -4 & 8 \\
   2 & -2 & 1 & 2 & 2 
   \end{pmatrix}
\smallebo
The overall minus sign is due to assumed anticommutation property of field operators. 
We also have the ``color Fierz'' identity
\smallbbo
\displaystyle\sum_{A} T^{Ai}{}_j T^{Ak}{}_{\ell} = \half\left( \delta^i_{\ell}\delta^k_{j}
- \frac{1}{N_c}\delta^i_{j}\delta^k_{\ell}  \right)
\smallebo
In particular we see from the second line of eq.(\ref{fmatrix}) that
\smallbbo
\backo \!\!\!\!\!
[\bar{\psi}_{L}\!\gamma_\mu\! T^A \psi_{L}]
[\bar{\psi}_{R}(y)\!\gamma_\nu T^A \psi_{R}]
=-[\bar{\psi}_{L}\psi_{R}][\bar{\psi}_{R}(y) \psi_{L}]
\smallebo
This yields to the  form of the interaction of eq.(\ref{coloronexchange}), to leading order in $1/N_c$,
\smallbbo
\label{coloronexchangeapp}
\backo\backo
S'=g_0^2\!\!\int_{xy}\!\! \; [\bar{\psi}_L(x)\psi_{R}(y)] D_F(x-y)[\bar{\psi}_R(y)\psi_{L}(x)],
\smallebo
If we now take the $\delta$-function limit of $D_F$
eq.(\ref{4NJL})
we have
 \smallbbo
\label{4NJLapp}
{D}_F(x-y)\rightarrow  \frac{1}{M_0^2}\delta^4(x-y),
 \smallebo\normalsize
and our interaction becomes the 4-fermion NJL interaction of eq.(\ref{NJL1}):
\smallbbo
\label{NJL1app}
\backo\backo
S'=\frac{g_0^2}{M_0^2}\!\!\int_{x}\!\! \; [\bar{\psi}_L(x)\psi_{R}(x)][\bar{\psi}_R(x)\psi_{L}(x)],
\smallebo
The ingredients of the Fierz rearrangement used here for Dirac matrices and color
can be found in the Appendix of ref.(\cite{HillThesis}).

\section{Notes on Scattering States  \label{Free}}

\subsection{Bilocal Scattering States}

 For subcritical coupling the SKG eigenvalue,  $\mu^2$, is positive and the large $r$ solution becomes
 $u(r)\sim a\exp(i\mu r) + b\exp(-i\mu  r)$.  This is a steady state sum of incoming
 and out going waves and represent the formation and decay of a resonance.
 This is an open scattering state description by a  bilocal field  of the resonance in
 the barycentric frame, $\Phi'(X,r)=\chi(X)u(r)/r$.
 
In the  weak coupling limit, $g_0<\!< g_{0c}$,  resonances appear in scattering states, 
centered at  positive invariant  (mass)$^2$, $\mu^2=\mu^2=k^2-g_0^2M_0^2$ with $\cos(\mu R_0)=0$.  
The full spatial solution for $\Phi'(X,r)$ will then be a wave-function with an 
an extended tail for large $r$,  consisting of incoming production 
and outgoing decay modes. As we increase the coupling $\mu^2$ decreases and approaches the scale invariant
critical value.

A scattering state is non-compact in $\vvr$ and technically non-normalizable due to the external spherical
scattering waves. For a narrow resonance
we can define an effective finite radius, $r\sim R_0$, of the bound state as a cut-off. 
For example, we might imagine something like a BEH boson composed of massless top quarks
in the symmetric phase of the standard model,
with a large positive $\mu^2$.  Such an object 
would therefore 
be a resonance in the $\bar{t}t$ scattering
amplitude with a width $\Gamma \propto \mu$.
The decay width can be estimated by computing the classical power 
in the outgoing wave divided by $\mu$.

As a simple example consider a rectangular (mass)$^2$ potential well,
 \smallbbo
 V_0(2r) = -g^2 M_0^2\theta(R_0 -r)
 \smallebo\normalsize
We focus upon s-wave scattering
and write the large $r$ the form 
\smallbbo
\backo\backo
\chi=e^{-i\mu t}, \;\; \phi(r) = \frac{u(r)}{r},\;\;
u(r)=\sin(kr+\delta)
\smallebo\normalsize
where $\delta $ is the phase shift.

For the attractive well 
$\mu^2=k'{}^2+V_0=k^2$,  $k'{}^2=\mu^2+g_0^2M_0^2$,
with
interior solution,
\smallbbo
\phi_{int}(r)=\frac{A\sin(k'r)}{r}   \qquad   k'{}^2+V_0=\mu^2
\smallebo\normalsize
and exterior solution,
\smallbbo
\phi_{int}(r)=\frac{A\sin(kr+\delta)}{r}   \qquad  k^2=\mu^2
\smallebo\normalsize
Matching:
\smallbbo
k'\cot(k'R_0)=k\cot(kR_0+\delta)
\smallebo\normalsize
hence,
\smallbbo
\backo
\tan\delta= \frac{
k\tan(k'R_0)-k'\tan(kR_0)}
{
k'+k\tan(kR_0)\tan(k'R_0)} 
\smallebo\normalsize
For small $kR_0<<1$ we have the scattering length,
\smallbbo
a_0\approx-\frac{\tan\delta_0}{k}\approx -\frac{\delta}{k}=-R_0\bl \frac{\tan{k'R_0}}{k'R_0} -1  \br
\smallebo\normalsize 
Total crossection, $\sim 4\pi \sin^2(\delta)/k^2$.

We can solve for $A$
\smbbo
\backo
|A|^2=\bl 1+ \left(\frac{g_0^2M_0^2}{\mu^2}\right)\cos^2(R_0\sqrt{\mu^2+g^2M_0^2})    \br^{-1}
\smallebo\normalsize
Resonances occur for maxima of $|A|^2$, these are approximately
the vanishing of $\cos(k'R_0)$ or $k'R_0=(n+1/2)\pi$.
We have $k'{}^2-g_0^2 M_0^2=\mu^2$.  

We can approximate the resonance by the lump contained within the potential,
which we normalize to unity as per our formalism for $\phi(r)$
This can then be use to compute Yukawa coupling $g_Y$.
We then have the decay width of a scalar particle of mass $\mu$ into
chiral fermions.
This yields,
\smallbbo
\Gamma = \frac{g_Y^2N_c}{16\pi} \mu   
\smallebo\normalsize

\subsection{General Notes and Kinematics}

For free particles of 4-momenta $p_{i\mu}$ we see that $\Phi$ describes a 
scattering state,
 \smallbbo
\Phi(x,y)=\exp(iP_\mu X^\mu + i Q_\mu r^\mu)
 \smallebo\normalsize
where,
 \smallbbo
P_\mu = p_{1\mu}+p_{2\mu} \qquad Q_\mu= p_{1\mu}-p_{2\mu}
 \smallebo\normalsize
For massive particles, $p_1^2=p_2^2=\mu^2$, and $\Phi$ then satisfies 
 \smallbbo
\label{dyn0}
\bl \frac{\partial^2}{\partial X^\mu\partial X_\mu} 
\!\!+\!\!\frac{\partial^2}{\partial r^\mu\partial r_\mu}\!\!+\!4\mu^2\br\Phi'(X,r) =0 
\smallebo
and the constraint is then,
\smallbbo
\label{cons}
\backo\backo\;\;\;\;
\frac{\partial^2}{\partial X^\mu\partial r_\mu}\Phi'(X,r) =0\rightarrow P_\mu Q^\mu=0
 \smallebo\normalsize
and  in the rest frame, $P_\mu=(2\mu,0,0,0)$,
hence $P_\mu Q^\mu=0$ implies $Q=(0,2\vvq)$.
Hence
$\Phi(X,\vvr)$ is independent of $r^0$ and the evolution
of the system is described by the single time variable, $X^0$.

In the rest frame we have,
 \smallbbo
Q^2=(p_1-p_2)^2=-(\vec{p}_1-\vec{p}_2)^2=-4\vec{q}\;{}^2
 \smallebo\normalsize
Hence, from eqs.(\ref{dyn0},\ref{cons})
 \smallbbo
 \frac{\partial^2}{\partial X^\mu\partial X_\mu}\Phi(X,\vvr)= 4(\mu^2+\vec{q}\;{}^2)\Phi(X,\vvr)
 \smallebo\normalsize
Therefore, $\Phi$  has continuum of
invariant ``masses'' $m_{\vvq\;{}^2}^2= 4(\mu^2+\vec{q}\;{}^2)$ 
This is  an ``unparticle,'' 
as in \cite{Georgi}. 

We factorize $\Phi=\chi(X)\phi(r)$ and the
factor field $\phi(r)$ then
satisfies the static SKG equation which generates the eigenvalue $4(\mu^2+\vec{q}\;{}^2)$,
 \smallbbo
-\nabla_{\vvr}^2\phi(\vvr) +4\mu^2\phi(\vvr)=4(\mu^2+\vec{q}\;{}^2)\phi(\vvr)
 \smallebo\normalsize
and the solutions of the factorized field $\phi(r)$ are static, box normalized, plane waves,
 \smallbbo
\phi(\vvr) = \frac{1}{\sqrt{V}}\exp (2i\vvq\cdot \vvr  )  
 \smallebo\normalsize
$\chi(X)$ then satisfies the KG equation with $X^0=t$ and $\vvX=0$,
 \smallbbo
\partial_t^2 \chi(t)+4(\mu^2+ {\vvq}\;{}^2)\chi(t)=0\qquad t=X^0
 \smallebo\normalsize

\subsection{More on the Removal of Relative Time \label{reltime}} 

 With secondary Lagrange multiplier constraints added to the action, we can define a timelike unit vector, $\omega^\mu$, 
 \smallbbo
 \backo
 \eta_1 \left|i\Phi^\dagger \frac{\partial}{\partial{X^\mu}} \Phi - P_\mu \Phi^\dagger \Phi\right|^2
 + \eta_2 \left|\omega_\mu \sqrt{P^\rho P_\rho}  -  P_\mu \right|^2
 \smallebo\normalsize
 We then define 
 \smallbbo
 Z_0=\delta(M\omega_\mu r^\mu).
 \smallebo\normalsize 
 The $\delta$-function removes 
 $\int dr^0$ in the rest frame but maintains manifest
 Lorentz invariance.
 
The normalizer, $Z_0$, is needed for composite fields. Fields are not
 directly observable, but their charges and current are. For the composite
 fields, which may describe bound states, we need charges that match those of the constituents 
 and, e.g., that count the number of bound states in a given quantum state.
 $Z_0$ normalizes these charges, seen  made more 
precisely below where we discuss the charges and currents in
the bilocal field theory, (see Appendix \ref{currentsapp} ,
for the discussion in the case of scalar field bilocal theory 
in \cite{bilocal}).  Note that $Z_0$ appears
only in the kinetic terms, where the currents arise,
and is not part of the interaction.

 \subsection{Comments on the Induced Yukawa Interaction \label{yukint}}

It is useful to consider the kinematics of the induced
Yukawa interaction.

In the rest
frame, suppose we have the decay of a resonant state of mass $\mu$ to 
a pair of free massless fermions.
Then $\chi(X^0)\sim e^{i\mu X^0}$, and 
 fermions with $\bar{\psi}_L(X\!+\!r)\sim \exp(ip_1(X\!+\!r))$
and $\psi_{R}(X\!-\!r)\sim \exp(ip_2(X\!-\!r))$.  The integral over $X$
yields energy and momentum conservation as usual:
\smallbbo
\backo
\mu = p_{10}+p_{20};\;\;\; 0=\vvp_{1}+\vvp_{2};\;\; \makebox{hence,}\;\; p_{10}=p_{20}
\smallebo\normalsize
We then have the integral over $r^0$ in eq.(\ref{Ypot1}),
\smallbbo
\backo\backo
\int\!\! d^4r\; e^{2i\vvp_{1}\cdot \vvr} D(2r)\phi(\vvr)
=-\half \int\!\! d^3r\; e^{2i\vvp_{1}\cdot \vvr} V_0(2\vvr)\phi(\vvr)
\smallebo\normalsize
Here we see that momentum conservation implies $p_{10}=p_{20}$,
hence $e^{i(p_{10}-p_{20})r^0}=1$, which allows us to integrate out
$r^0$ over $D(2r)$ with the static $\phi(\vvr)$ as before. 
The  remaining integral over $2\vvr$ is  the Fourier transform of
$V_0(2\vvr)\phi(\vvr) $  with the 3-momentum difference $\vvp_1-\vvp_2=2\vvp_{1}$ flowing through
the extended vertex.  
If the mass scale associated with $V_0(2\vvr)\phi(\vvr) $
is large, i.e. $M_0^2>\!\!>\mu^2$, then we can reliably replace this with,
\smallbbo
\backo\backo
\!\!\int\!\! d^3r\; e^{2i\vvp_{1}\cdot\vvr}V_0(2\vvr)\phi(\vvr)\sim \!\!\int\!\! d^3r\; V_0(2\vvr)\phi(\vvr)+{\cal{O}}\bl\!
\frac{\vvp_{1}{}^2}{M_0^2}\!\br
\smallebo\normalsize
An interesting feature is that the decay amplitude of a resonance depends only
upon the part of the wave-function localized within the potential. 

On the other hand we can suppose that the field $\chi$ has developed a VEV 
$\langle \chi \rangle = f_0$
(which is
the case with supercritical coupling as we discuss below).
Then the internal wave-function, $\phi(\vvr)$, is a localized solution
of the SKG equation.
In this case the free fermion mass is spontaneously generated by the Yukawa coupling
to the VEV.

In this case we have fermions of zero 3-momentum,  an 
incoming fermion, $\sim \exp(ip_1(X\!+\!r))$ and outgoing fermion
$\sim \exp(-ip_2(X\!-\!r))$. Now the $d^4X$ integral yields $p_1^0=p_2^0=\mu$,
where $\mu$ is the induced fermion mass,
and 
\smallbbo
\backo 
\int\!\! d^4r \; e^{2i\mu r^0}D(2r)\phi(\vvr)
=-\half \int\!\! d^3r \; V'_0(2\vvr,\mu)\phi(\vvr)
\smallebo\normalsize
where now the potential is slightly distorted,
\smallbbo
\backo 
V'_0(2\vvr,\mu) =-\frac{M'e^{-2M'|\vvr|}}{8\pi |\vvr|}\qquad M'{}^2=M_0^2-\mu^2
\smallebo
Assuming $\mu <\!\!<M_0$ we have, 
$V'_0(2\vvr,\mu)\approx V_0(2\vvr) +{\cal{O}}(\mu^2/M_0^2)$.

\vspace{0.1in}

\section{Summary of Notation $\label{SumNote}$}

Barycentric coordinates:
 \smallbbo
\label{barycentric}
X=\half(x+y)\qquad \rho = (x-y)\qquad r=\half(x-y)
\nonumbo
\partial_x=\half(\partial_X+\partial_r) 
\;\;\;
\partial_y=\half(\partial_X-\partial_r)
 \smallebo
Integrals:
 \smallbbo \!\!\!
\int_{u...v}\!\!\!\! =\int\!\!d^4u..d^4v;\qquad \int_{\vvx...\vvy}\!\!\!\! =\int\!\!d^3x..d^3y
\nonumbo \!\!\!
 \int_{u...v;\vvx...\vvy}\!\!\! =\int\!\!d^4u..d^4v\;d^3x...d^3y;
\;\;\;
\int \!\!d^4xd^4y =J\!\! \int d^4X d^4r 
\nonumbo \!\!\!\!
\makebox{Jacobian, $J=(2)^4$:}
\;\;
\int\!\! d^4xd^4y\bl|\partial_x\phi|^2+|\partial_y\phi|^2-\mu^2|\phi|^2\br
\nonumbo = J\!\!\int d^4Xd^4r\bl\half|\partial_X\phi|^2+\half|\partial_r\phi|^2-\mu^2|\phi|^2
\br 
\smallebo


\section*{Acknowledgments}
I thank  Bill Bardeen, Bogdan Dobrescu and  Julius Kuti for
discussions over the long duration of this project.  I also thank
the  Fermi Research Alliance, LLC under Contract No.~DE-AC02-07CH11359 
with the U.S.~Department of Energy, 
Office of Science, Office of High Energy Physics,
and The University of Wisconsin Physics Department for an Honorary Fellowship.

\end{document}